\documentclass[journal=jpcbfk,manuscript=article]{achemso}

\usepackage[version=3]{mhchem} 
\usepackage{natbib}
\usepackage{xcolor}

\newcommand*{\p}[1]{\left(#1\right)}

\author{Huanghao Mai}
\affiliation[Caltech]
{Division of Chemistry and Chemical Engineering, California Institute of Technology}
\email{hmmai@caltech.edu}
\author{Matthew H. Zimmer}
\affiliation[Caltech]
{Division of Chemistry and Chemical Engineering, California Institute of Technology}
\author{Thomas F. Miller III}
\affiliation[Caltech]
{Division of Chemistry and Chemical Engineering, California Institute of Technology}

\title[]
  {Exploring PROTAC cooperativity with coarse-grained alchemical methods}

\begin{document}

\begin{tocentry}

\includegraphics[page=1,scale=0.55]{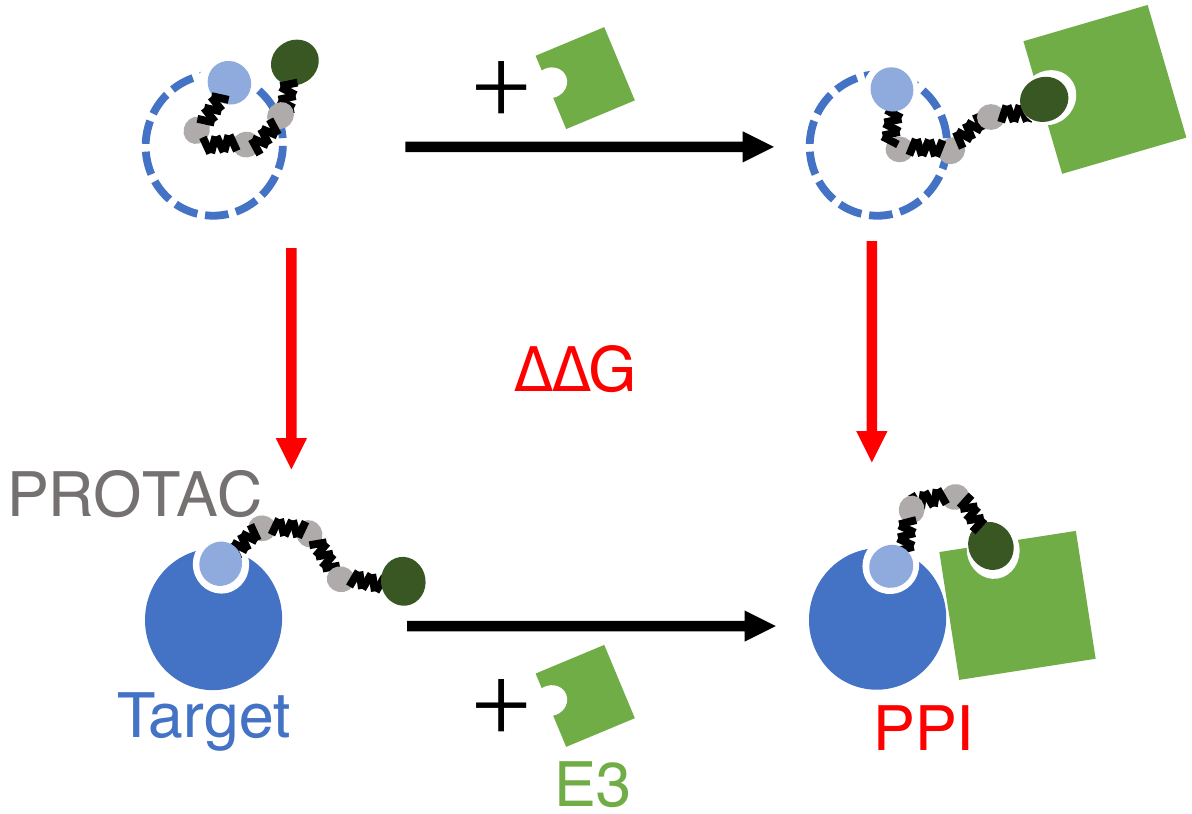}

\end{tocentry}

\begin{abstract}
Proteolysis targeting chimera (PROTAC) is a novel drug modality that facilitates the degradation of a target protein by inducing proximity with an E3 ligase. In this work, we present a new computational framework to model 
\textcolor{black}{
the cooperativity between PROTAC-E3 binding and PROTAC-target binding  
} 
principally through protein-protein interactions (PPIs) induced by the PROTAC. Due to the scarcity and low resolution of experimental measurements, the physical and chemical drivers of these non-native PPIs remain to be elucidated. We develop a coarse-grained (CG)
approach to 
model interactions in the target-PROTAC-E3 complexes, 
which enables converged thermodynamic estimations using alchemical free energy calculation methods despite an unconventional scale of perturbations. With minimal parameterization, 
we successfully capture 
fundamental principles of cooperativity, including the optimality of intermediate PROTAC linker lengths that originates from configurational entropy. 
We qualitatively characterize the dependency of cooperativity on PROTAC linker lengths and protein charges and shapes. Minimal inclusion of sequence- and conformation-specific features in our current forcefield, however, limits quantitative modeling to reproduce experimental measurements, but further development 
of the CG model may allow for efficient computational screening to optimize PROTAC cooperativity.
\end{abstract}

\section{Introduction}
\textcolor{black}{Proteolysis targeting chimera (PROTAC)} has emerged as a promising drug modality that elicits protein degradation by hijacking the ubiquitin-proteasome system (UPS), a major regulatory component of cells. In the UPS pathway, E3 ligases transfer ubiquitins onto aberrant proteins to mark them for degradation by proteasomes. A PROTAC molecule exploits this pathway with two binding moieties that tether the target protein and an E3 ligase together. The tethered target protein thus becomes a neo-substrate of the E3 ligase and is subsequently ubiquitinated for proteasomal degradation. 
PROTACs require a lower dose than conventional small-molecule inhibitors because of their catalytic nature and they have the potential to target the undruggable proteome \cite{an_small-molecule_2018,burslem_proteolysis-targeting_2020}.
Since the first proof-of-concept in 2001 \cite{sakamoto_protacs_2001}, the number of proteins successfully degraded by PROTACs has grown rapidly, and examples \textcolor{black}{of such proteins} include kinases and gene regulators that are implicated in cancer. As of 2021, at least 13 PROTACs are in or approaching clinical trials \cite{mullard_targeted_2021}.

Despite increasing applications, there is a lack of guidance on designing PROTACs due to the unique mode of action \cite{maniaci_bifunctional_2019,troup_current_2020,alabi_major_2021}. 
In particular, a critical step in the degradation process is the formation of the ternary complex of target-PROTAC-E3. The ternary complex involves molecular interactions beyond the binary bindings between the two warheads of a PROTAC and the two proteins. The selectivity \cite{gadd_structural_2017,nowak_plasticity_2018,smith_differential_2019} and stability \cite{riching_quantitative_2018,roy_spr-measured_2019,farnaby_baf_2019,du_structural_2019} of the ternary complex can both be improved through favorable 
\textcolor{black}{
protein-protein interactions (PPIs)}
between the target protein and the E3 ligase. 
For certain targets, the degradation outcome can be very different depending on whether cereblon (CRBN) or von Hippel-Lindau (VHL), the two most heavily used E3 ligases, more efficiently and selectively form a productive complex with the target
\cite{lai_modular_2016,bondeson_lessons_2018,riching_quantitative_2018,donovan_mapping_2020}. As more warheads for E3 ligases are designed \cite{spradlin_harnessing_2019,ward_covalent_2019,zhang_electrophilic_2019,kuljanin_reimagining_2021}, choosing which of the more than $600$ E3 ligases in humans \cite{li_genome-wide_2008} optimally interact with the target protein will become important \cite{jevtic_e3_2021,scholes_identification_2021}. While PPIs depend on the sequences and the structures of the proteins, PROTACs can also modulate the PPIs by restricting the distance and relative orientation between the target and the E3 ligase, effectively changing the entropic component of PPIs. 

Because of this three-body interplay and the transient nature of the ternary complex, a complete characterization of the PPIs as a function of the PROTAC, the target protein, and the E3 ligase is intractable. A few proteomics studies \cite{bondeson_lessons_2018,huang_chemoproteomic_2018,donovan_mapping_2020} on kinase degradation have used PROTACs with promiscuous warheads such that the PROTAC-induced PPIs differentially affect the degradation outcome of hundreds of proteins. These studies reported the fold change of protein abundance due to PROTAC treatment, but analysis can be complicated by secondary interactions \cite{scholes_identification_2021} 
\textcolor{black}{
and numerous other factors such as the permeability of the PROTAC, half-lives of the target proteins, cellular localization, and reactions downstream of ternary complex formation \cite{rodriguez-rivera_unifying_2021}. Other studies \cite{gadd_structural_2017,maniaci_homo-protacs_2017,nowak_plasticity_2018,chan_impact_2018,zorba_delineating_2018,schiemer_snapshots_2021}}
have focused on specific target-E3 pairs and examined the effect of changing PROTAC properties such as the linker length. They measured the difference in the strength of PROTACs binding to the target or the E3 ligase due to the presence of the other protein. This difference, termed binding cooperativity, reflects the strength of PROTAC-mediated PPIs. 
However, few generalizable patterns have emerged and systematic experimental characterizations remain scarce.

Computational modeling based on docking or atomistic molecular dynamics (MD) has complemented experimental work  \cite{nowak_plasticity_2018,zorba_delineating_2018} and displayed promising future prospects, but there are several limitations to current methodologies. Although standard docking protocols don't handle three-body problems, several workflows have been adapted ad hoc for PROTAC \cite{drummond_improved_2020,zaidman_prosettac_2020,weng_integrative_2021,bai_rationalizing_2021,bai_modeling_2022}. Docking studies rank ternary complex conformations by scoring functions biased for naturally evolved PPIs and benchmark against the few crystal structures of PROTAC-induced ternary complexes \cite{moreira_proteinprotein_2010,gromiha_proteinprotein_2017,bemis_unraveling_2021}. The results can be inaccurate as PROTAC-induced PPIs are non-native and exhibit plasticity \cite{nowak_plasticity_2018,eron_structural_2021}. 
In contrast, atomistic MD is physically grounded to capture non-native PPIs. However, the size of the ternary complex modeled at an atomistic resolution significantly limits the timescale of simulations, such that naively simulating PPIs can be prohibitively slow. Sophisticated enhanced sampling techniques and distributed computing are needed to sample an ensemble of low-energy conformations that are consistent with experimental data \cite{dixon_atomic-resolution_2021}. 
Due to the difficulties in modeling the ternary complex, direct calculation of the binding cooperativities \textcolor{black}{was} not attempted until two recent studies \cite{li_importance_2022,liao_silico_2022} that explored the molecular mechanics with the generalized Born and surface area continuum solvation (MM/GBSA).

Here, we seek an orthogonal approach that combines \textcolor{black}{coarse-grained MD (CGMD)} and alchemical free energy calculation methods to study PROTAC cooperativities.  
On the spectrum of computational tools, docking and atomistic MD are positioned at the empirical and first-principle ends respectively, and finding a compromise in the middle of this spectrum is a promising direction.
Compared to atomistic modeling, coarse-graining reduces the effective size of the model and smoothens the energy surface, enabling simulations at a much longer timescale necessary for the PROTAC-mediated complexes. 
While CGMD may struggle to recapitulate the molecular basis of lock-and-key bindings, such strong and specific interactions are less imperative in non-native PPIs induced by PROTACs. Moreover, PROTAC binding reduces the ways proteins can interact with each other, differentiating and simplifying the problem studied here from the formidable task of modeling general protein-protein binding. 
In docking, such constraints are challenging to incorporate into the scoring functions and are approximated through separate steps to filter compatible PPI poses and PROTAC geometries. \textcolor{black}{While CGMD excludes many degrees of freedom from the PROTAC, proteins, and solvent entropy, this effect of  configurational entropy on PPIs from PROTAC mediation can be directly captured. }
Finally, we calculate binding energies using alchemical methods, which circumvents the computational challenge of directly sampling binding and unbinding events between the PROTAC and proteins. We demonstrate the computational amenity of an unconventional application of alchemical methods motivated by the PROTAC systems, 
and take advantage of the physical interpretability of the CGMD + alchemical approach to explore the principles of PROTAC binding cooperativity.

\section{Methods}

\subsection{CGMD setup of PROTAC-protein complexes}

The binary and ternary PROTAC-protein complexes are coarse-grained at two resolutions to efficiently sample complex conformational changes while retaining sufficient details for structural insight. Specifically, a major focus of this work is to characterize the entropic effect of the length of PROTACs on the strength of induced PPIs, necessitating modeling the PROTAC linker at a higher resolution than the rest of the system.
Proteins are coarse-grained by mapping every three amino acids onto a large bead of $\sigma=0.8$ nm diameter, which is approximately the Kuhn length of polypeptides \cite{niesen_structurally_2017,zhang_long-timescale_2012,hanke_stretching_2010,staple_model_2008}. Binding moieties at the two ends of a PROTAC are each represented by a large bead, whereas the linker region is modeled as a Gaussian chain at the resolution of a PEG unit ($\sigma_s=0.35$ nm \cite{charmainne_cruje_polyethylene_2014}) or three heavy atoms. 
Several experimental works that used flexible linear linkers motivate our modeling approach for the PROTAC linker, including Chan et al. \cite{chan_impact_2018} where an alkane linker was varied in step sizes of our linker beads and Zorba et al. \cite{zorba_delineating_2018} where a PEG linker is modified at smaller length steps such that linker lengths ranging from 1 to 6 $\sigma_s$ in our modeling correspond to the PROTAC (1), (3), (5), (6), (8), and (10).

A minimal forcefield is used to describe the internal and interactive forces, and a full description can be found in the Supporting Information (Section S1). The three-dimensional structure of a protein is maintained by a bottom-up fitted elastic network model (Fig. S2), which allows conformational flexibility \cite{lezon_elastic_2009,ricardobatista_consensus_2010}. Protein beads can have additional properties to describe PPIs beyond volume exclusion (Fig. S1). When modeling electrostatic interactions, for example, a protein bead has the net charges of the triplet of residues that it is coarse-grained from. PROTACs are modeled as Gaussian polymers with volume exclusion, and the warhead beads are attached to the binding pockets of proteins through harmonic springs. Modeling PROTAC interactions beyond warhead binding is out of the scope of this work. Thus, under current setup, PROTAC beads have 0 charge and no affinity to any other beads.

\begin{figure}[H]
\centering
\includegraphics[page=1,scale=0.48]{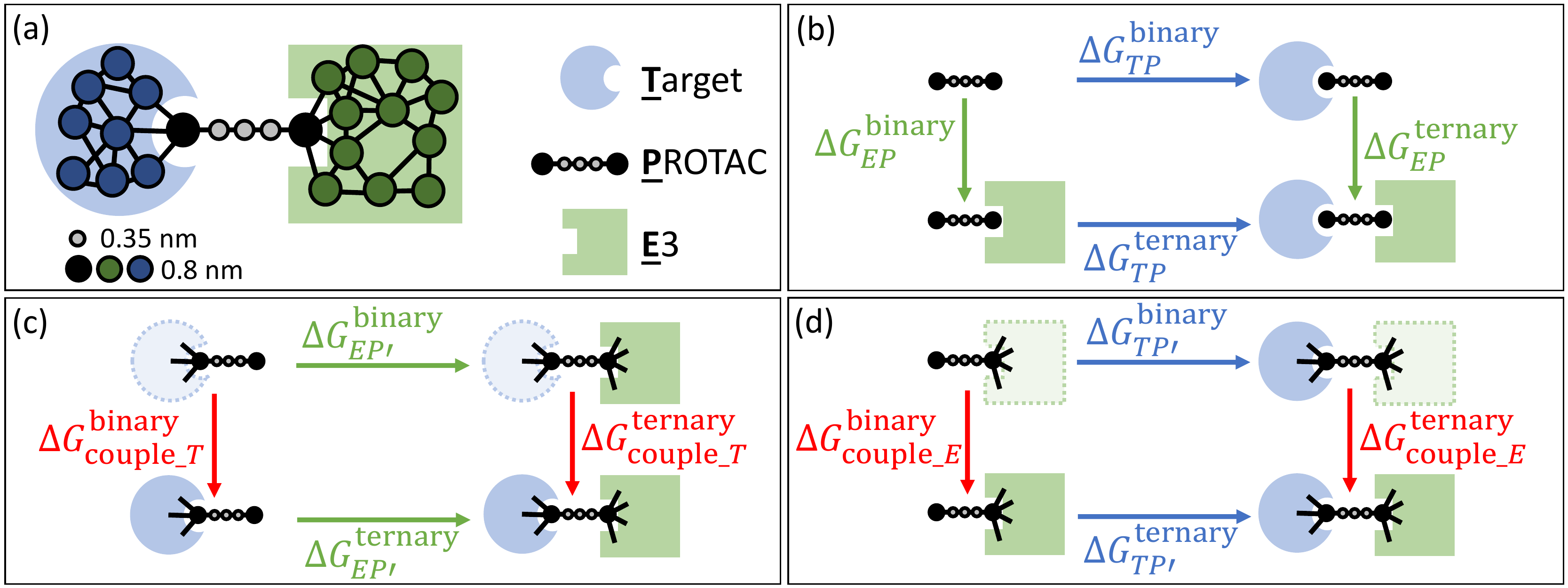}
\caption{Schematic of the simulation setup for PROTAC-mediated complexes. \textbf{(a)} The target-PROTAC-E3 ternary complex is initialized with a fully extended PROTAC as drawn. The proteins are coarse-grained at the resolution of three amino acids per bead, approximately $0.8$ nm. PROTAC warhead beads are represented by beads of the same size, whereas the linker is coarse-grained at a higher resolution. 
\textbf{(b)} The thermodynamic cycle shows that the $\Delta \Delta G$ is obtained by measuring the free energy difference between target-PROTAC binding with and without the E3 ($\Delta G^\text{binary}_{TP}-\Delta G^\text{ternary}_{TP}$) or between PROTAC-E3 binding with and without the target ($\Delta G^\text{binary}_{EP}-\Delta G^\text{ternary}_{EP}$). Under the alchemical setup, $\Delta \Delta G$ can be alternatively quantified by the free energy difference between the red vertical processes, which represent coupling the target ($\Delta G^\text{binary}_{\text{couple\_}T}-\Delta G^\text{ternary}_{\text{couple\_}T}$ in \textbf{(c)}) or the E3 ($\Delta G^\text{binary}_{\text{couple\_}E}-\Delta G^\text{ternary}_{\text{couple\_}E}$ in \textbf{(d)}) to the PROTAC and the PROTAC pre-bound to the other protein. In the initial states in \textbf{(c)} and \textbf{(d)}, \textcolor{black}{the dotted lines represent the target or the E3 whose interactions with the rest of the system are turned off except for the harmonic constraints (black lines) to the PROTAC warhead.}}
\label{fgr:geo_cycle}
\end{figure}

The orientation between the E3 ligase and the target protein is initialized such that the two binding pockets face each other, with a fully extended PROTAC tethering in between (Fig. \ref{fgr:geo_cycle}a). The binding moiety beads of PROTAC are placed at the center of each binding pocket, which is defined by the residues within 4 or 5 Å from the PROTAC warhead in experimental structures. Thus, setting up the initial coordinates of a ternary complex requires the following inputs: structures of each protein, residues at the two PROTAC binding pockets, and the length of the PROTAC linker. To calculate the difference in PROTAC binding energies due to PPIs, simulations of binary target/E3-PROTAC complexes are also needed. Binary complexes are prepared by removing a protein from the initialized ternary complex. 

\subsection{Thermodynamic framework of alchemical perturbation} 

The binding cooperativity of a PROTAC is mathematically defined as $\exp{\p{\frac{\Delta \Delta G}{RT}}}$, where $R$ is the gas constant, $T$ here refers to the temperature in the context of an energetic scale and refers to the target protein elsewhere, $\Delta \Delta G = \Delta G^\text{binary}_{TP}-\Delta G^\text{ternary}_{TP}$, and $\Delta G^\text{ternary}_{TP}$ and $\Delta G^\text{binary}_{TP}$ are the free energies of the PROTAC ($P$) binding to the target protein ($T$) with and without the presence of the E3 ligase ($E$).
Because of the thermodynamic cycle (Fig. \ref{fgr:geo_cycle}b), the same $\Delta \Delta G$ can be obtained from $\Delta G^\text{binary}_{EP}-\Delta G^\text{ternary}_{EP}$. Favorable PPIs stabilize the ternary complex and facilitate PROTAC binding to both proteins. Thus, they lower $\Delta G^\text{ternary}_{TP}$ and $\Delta G^\text{ternary}_{EP}$, which leads to larger $\Delta \Delta G$ and more positive cooperativity. 

Alchemical free energy calculation methods exploit alternative thermodynamic cycles to obtain $\Delta \Delta G$ without simulating binding and unbinding processes. For simplicity, in this work, all $\Delta \Delta G$s are calculated using the cycle in Fig. \ref{fgr:geo_cycle}c, which we describe in detail here, but one should arrive at the same result using the mirroring cycle in Fig. \ref{fgr:geo_cycle}d. 
By the definition of a thermodynamic cycle, we have $\Delta G^\text{binary}_{EP'}-\Delta G^\text{ternary}_{EP'} = 
 \Delta G^\text{binary}_{\text{couple\_}T}-\Delta G^\text{ternary}_{\text{couple\_}T}$,
where $\Delta G^\text{binary}_{\text{couple\_}T}$ and $\Delta G^\text{ternary}_{\text{couple\_}T}$ represent the free energies of coupling $T$ to $P$ and to the target-PROTAC bound complex $EP$. In the initial states of both coupling processes (vertical processes in red in Fig. \ref{fgr:geo_cycle}c), $T$ is bound to $P$ but is a dummy molecule at an ideal state. Specifically, multiple harmonic springs connect the binding pocket beads in $T$ to the warhead bead of $P$, and $T$ itself is an elastic network model consisting of only harmonic springs. All other interactions between $T$ and the rest of the system -- whether $P$ or $EP$ -- are turned off. \textcolor{black}{Coupling $T$} simply means turning on these inter-molecular interactions, while \textcolor{black}{the binding pocket springs remain unperturbed.}

\textcolor{black}{Attaching a dummy $T$ instead of having $T$ dissociated results in a systematic error in the horizontal free energies of $EP$ binding ($\Delta G^\text{binary}_{EP'}$ and $\Delta G^\text{ternary}_{EP'}$ in Fig. \ref{fgr:geo_cycle}c) such that the $\Delta \Delta G$ is unaffected. This is because the attachment of dummy $T$ occurs via only one bead on $P$}, except which there are no other forcefield terms involving both physically present beads and dummy beads. In the configurational partition function, energy terms describing the geometries of the physically present part of the system can therefore be separated from the term involving the dummy $T$ and the attachment junction. The latter term is the same whether the physically present part is $P$ or $EP$, such that the unphysical contribution from attaching dummy $T$ cancels out in $\Delta \Delta G$.  

\subsection*{Free energy calculations}
\label{FEP_calc_section}
Alchemically creating a protein from a dummy state to \textcolor{black}{full coupling} involves turning on the interaction potentials between the protein and the rest of the system in the forcefield. The interactions are turned on \textcolor{black}{in stages by sequentially scaling each kind of interaction potential using a coupling parameter $\lambda$. Intramolecular potentials (e.g. the elastic network model of each protein) and intermolecular potentials not perturbed at the current stage are unaffected by the $\lambda$ scaling.} For the electrostatic potential, the start state (no electrostatics) and the end state (full electrostatics) correspond to $\lambda_\text{elec}=0$ and $1$ respectively. Intermediate states are interpolated 
such that the potential is defined as 
$U_{\lambda_\text{elec}} = (1-\lambda_\text{elec})U_\text{no\_elec} + \lambda_\text{elec}U_\text{elec} = \lambda_\text{elec}U_\text{elec}$. For numerical stability, the electrostatic potential is only perturbed in the presence of volume exclusion \cite{pohorille_good_2010,klimovich_guidelines_2015}, which is modeled by Weeks-Chandler-Andersen (WCA) potential.
To turn on Lennard-Jones (LJ) or variants of LJ potentials (e.g. WCA), a soft-core scaling \cite{mey_best_2020} with $\lambda_\text{LJ}$ is used for numerical stability: 

\begin{equation*}
U_{\lambda_\text{LJ}}\p{r_{ij}} = 4\epsilon \lambda_\text{LJ} \p{
                \frac{1}{\p{\alpha\p{1-\lambda_\text{LJ}}+\p{\frac{r_{ij}}{\sigma_{ij}}}^6}^2}-
                \frac{1}{\alpha\p{1-\lambda_\text{LJ}}+\p{\frac{r_{ij}}{\sigma_{ij}}}^6}
},
\end{equation*}

where $\alpha=0.5$, $r_{ij}$ is the distance between beads $i$ and $j$, and $\sigma_{ij}$ is the sum of the radii of beads $i$ and $j$. The number of intermediate states and the spacing of the coupling parameter values depend on the difficulty to obtain converged free energy calculations. For the electrostatic potential, a linear pathway where $\lambda_\text{elec}$ ranges from 0 to 1 with a step size of 0.125 is a simple and effective approach. For LJ and related potentials, because most of the free energy changes occur near the start state of $\lambda_\text{LJ}=0$ (Fig. \ref{fgr:convergence}b,c), we introduce intermediate states at $\lambda_\text{LJ}=0.005, 0.01, 0.015, 0.02, 0.04, 0.06, 0.08, 0.1, 0.2, 0.3, 0.5, 0.7,$ and $0.9$.

The $\Delta G$ of turning on each kind of interaction is calculated using thermodynamic integration (TI) \cite{kirkwood_statistical_1935}, Bennett acceptance ratio (BAR) method \cite{bennett_efficient_1976} and the multi-state BAR (MBAR) method \cite{shirts_statistically_2008}. 
TI and BAR/MBAR are distinct formulations for free energy calculations, and we verify that these methods converge to similar values. 
The system in CGMD is evolved using overdamped Langevin dynamics with a diffusion coefficient of 253 $\text{nm}^2\mathbin{/}$s and a timestep of 30 ns for stable integration.
At each state, at least 64 trajectories of 6 s long are generated to sample the conformations of the complexes. After collecting the samples from trajectories, post-processing involves calculating $\frac{\partial U}{\partial \lambda}$ and $\Delta U_{ij}$ for all $i,j=1,2,...,K$ states as inputs for TI, BAR, and MBAR.

\section{Results and discussion}

\subsection{Alchemical perturbation of protein domains is feasible with CGMD}
The binding cooperativity of PROTAC due to PPIs is a unique challenge that calls for an unconventional application of alchemical free energy calculation methods. Alchemical methods are mainly used to determine the binding energies between small-molecule ligands and proteins, and typically no more than 10 heavy atoms are perturbed for efficient and accurate calculations. In protein-protein binding, recent applications and development focus on quantifying the relative free energy changes from small-scale perturbations such as mutations of single residues \cite{clark_free_2017,clark_relative_2019,patel_implementing_2021,la_serra_alchemical_2022,nandigrami_computational_2022}. To our knowledge, the only case that alchemically calculates PPIs in a three-body setting compares how analogs of inhibitors change aberrant multimerization of the HIV-1 integrase \cite{sun_computational_2021}. Their proposed thermodynamic framework involves calculating the relative free energy difference by perturbing small molecules that directly participate at a fixed PPI interface. This framework is more readily extendable to molecular glues that modulate PPIs in a similar way. PROTACs, however, due to a more modular design, are typically larger linear molecules. The flexibility of the linker is often nontrivial, such that the two proteins cannot be kept bound at a fixed interface. This configurational entropic concern necessitates an unusually large perturbation at the scale of a protein rather than a small molecule to calculate the binding cooperativity, testing the computational limit of alchemical methods.    

\begin{figure}[htbp!]
\centering
\includegraphics[page=2,scale=0.48]{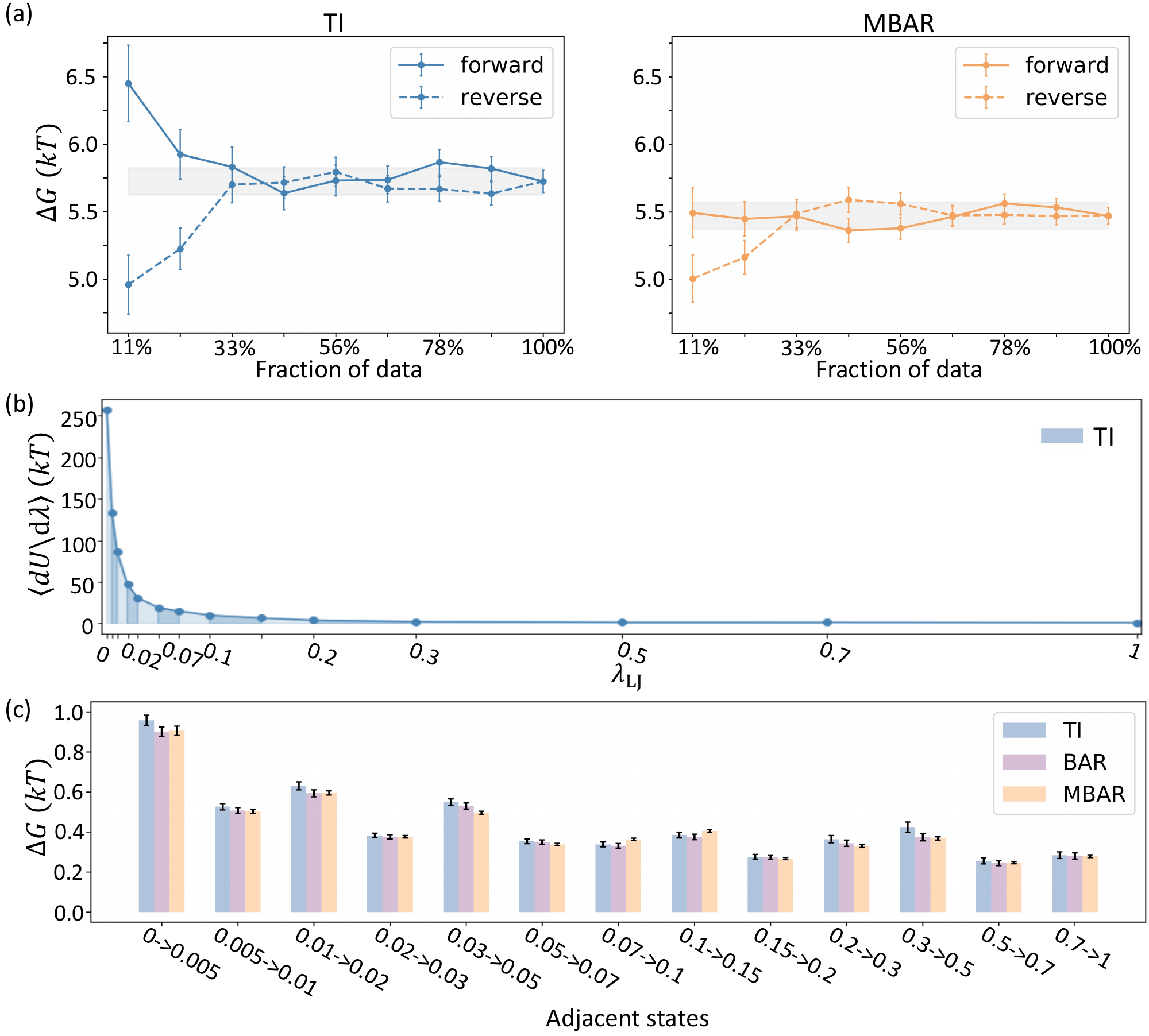}
\caption{Calculation of $\Delta G\textsuperscript{ternary(sterics)}$ by alchemical perturbation of BTK in the ternary complex of BTK-PROTAC (10)-CRBN.
\textbf{(a)} TI and MBAR both reach apparent convergence in the time-forward and time-reversed directions with no pathological signs. The grey band in each panel represents the final estimation using $100\%$ data $\pm 0.1$ $kT$ as a threshold for error tolerance, where $k$ is the Boltzmann constant.
\textbf{(b)} TI estimation is shown as the blue area under the curve of $\langle \partial U/\partial \lambda \rangle$. 
\textbf{(c)} TI, BAR, and MBAR agree for all intermediate $\Delta G$s between adjacent states. All error bars of computational results here and in subsequent figures represent $\pm 1$ std. Color coding for TI, BAR, and MBAR results are the same in subsequent figures unless otherwise stated.}
\label{fgr:convergence}
\end{figure}

To explore the feasibility of the CG alchemical approach, we calculate the free energy of turning on the steric repulsions between a target protein and a PROTAC-E3 complex ($\Delta G\textsuperscript{ternary(sterics)}$) in the absence of other inter-molecular potentials. We choose Bruton’s tyrosine kinase (BTK) as the target (only the kinase domain modeled), CRBN as the E3, and the PROTAC (10) from \cite{zorba_delineating_2018}, which are respectively modeled by 87, 124, and 8 beads in the CG model. Together they form the largest target-PROTAC-E3 complex simulated in this work.
We compare the calculations using different percentages of the simulation data collected in the time-forward and time-reversed directions. The \textcolor{black}{calculated values of $\Delta G\textsuperscript{ternary(sterics)}$} plateau starting around the midpoint of the simulation time, indicating numerical convergence (Fig. \ref{fgr:convergence}a). The time-forward and -reversed estimations are within 1 standard deviation (std) at the midpoint, and the time-reversed estimations remain stable after the midpoint. The observed behavior of the estimates over time suggests that unequilibrated samples at the beginning of the trajectories have been removed, and the remaining frames sample from similar distributions rather than distinct metastable states with slow transition rates \cite{klimovich_guidelines_2015}. 

Three methods, TI, BAR, and MBAR are used to separately estimate the free energies. The accuracy of all three methods depends on the number and the spacing of alchemical states. BAR and MBAR reweight conformations sampled from one state by their probability in another state to estimate the free energy differences. Having similar probability distributions between states, i.e. phase space overlap, is therefore critical to the estimation. 
Unlike BAR/MBAR, TI estimates the free energies by numerically integrating $\langle \frac{\partial U}{\partial \lambda} \rangle$, the ensemble average of the derivative of the potential energy $U$ along the alchemical pathway defined by $\lambda$. Depending on the curvature of $\langle \frac{\partial U}{\partial \lambda} \rangle$, choices of intermediate states specified by $\lambda$ and the integration scheme together introduce integration errors in addition to the statistical errors in estimating the ensemble average per state. 
We choose an alchemical pathway that involves 12 intermediate states in addition to the start and end states, such that $\Delta G\textsuperscript{ternary(sterics)}=\sum_{i=1}^{13} \Delta G_{\lambda_i,\lambda_{i+1}}$, where $\Delta G_{\lambda_i,\lambda_{i+1}}$ is the free energy of changing the WCA potential between neighboring states $\lambda_i$ and $\lambda_{i+1}$. With a total of 14 states unevenly spaced, the phase space overlap between neighboring states is sufficient (Fig. S3) for efficient reweighting-based estimations.  
For TI, the trapezoid rule of numerical integration is used for its simplicity and robustness. Although the quadrature errors result in a slight overestimation of $\Delta G\textsuperscript{ternary(sterics)}$, the $\partial U/\partial \lambda$ curve is sufficiently smooth such that TI and MBAR largely agree.
In addition to the global agreement on $\Delta G\textsuperscript{ternary(sterics)}$, TI, BAR, and MBAR also locally agree with each other on all $\Delta G_{\lambda_i,\lambda_{i+1}}$ along the alchemical pathway (Fig. \ref{fgr:convergence}c). 
We emphasize that TI and BAR/MBAR rely on distinct types of input data and processing procedures, and their consistency even at the most granular level of calculations further validate our CG alchemical approach. 

Analyses of estimations over simulation time and using different free energy calculation methods indicate that convergence of perturbing a protein can be achieved within reasonable computation time, significantly pushing the boundaries of applying alchemical methods. As parallelization can be done over the alchemical states and over trajectories for each state, the time to run one trajectory is the main limiting factor in the wall-clock computation time of applying our method. Criteria to determine how long a trajectory should be run are described in the Supporting Information (Section S2). 
For this work, depending on the size of the system, 3 - 14 CPU hours per trajectory of ternary complexes are sufficient.

\subsection{Minimal forcefield captures entropic effects in PROTAC-mediated PPIs}

Encouraged by the proof-of-concept calculations 
above for $\Delta G\textsuperscript{ternary}$, we also calculate $\Delta G\textsuperscript{binary}$ and complete our calculations for the $\Delta \Delta G$ of the thermodynamic cycle. \textcolor{black}{We follow the sign convention of $\Delta \Delta G$ such that a positive value represents positive cooperativity.} The BTK-CRBN system modeled here has been experimentally shown to lack large cooperativity, and 
introducing PROTACs in Hydrogen/Deuterium Exchange experiments didn't reveal significant profile changes that would indicate the presence of stable PPIs.
As the starting point for our method development, we focus on this system due to its apparent simplicity and the availability of experimental characterization over a large range of PROTAC linker lengths. We characterize $\Delta \Delta G$ changes over PROTAC lengths because this relies on capturing the fundamental physics of the tertiary interactions rather than sequence- or conformation-specific properties. 

Two forcefield setups are used to describe PPIs and the resulted $\Delta \Delta G$ trends over PROTAC linker lengths are compared.
In the first setup, we calculate the baseline $\Delta \Delta G$ in the absence of PPIs other than volume exclusion. In the second setup, nonspecific attractions between BTK and CRBN beads are added and explored at two strengths.
 The intrinsic PPIs without PROTAC mediation should be weak such that in the limit of infinite linker length the $\Delta \Delta G$ is negligible. The attenuation of weak PPIs with increasing PROTAC linker lengths originates from configurational entropy. As the PROTAC becomes longer, it experiences a greater loss of configurational freedom upon binding to proteins to induce PPIs, incurring an entropic cost. We examine this configurational entropic effect by modeling $\Delta \Delta G$ at linkers ranging from 1 to 6 beads ($\sigma_s$) long, which correspond to approximately 3.5 Å to 21 Å.

\begin{figure}[htbp!]
\centering
\includegraphics[page=3,scale=0.48]{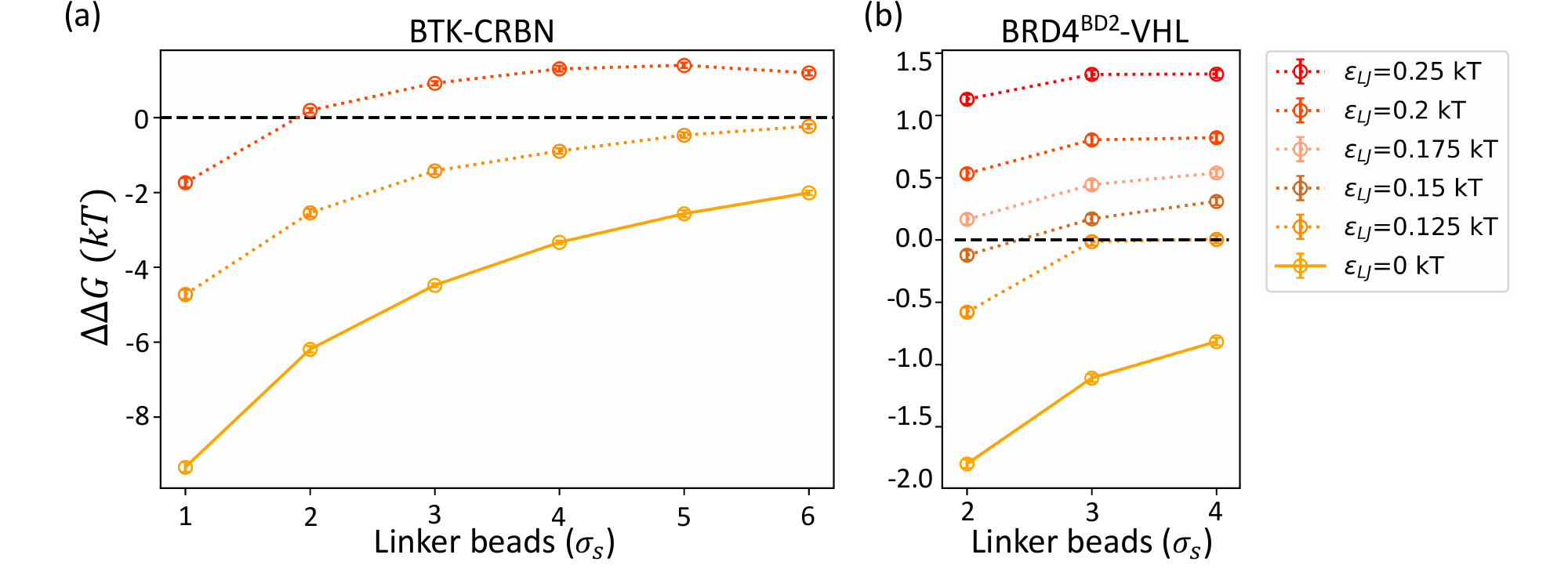}
\caption{The \textcolor{black}{$\Delta \Delta G$} trends over PROTAC linker lengths change with varying strengths of non-specific attraction between proteins. All calculations shown are obtained using MBAR, and results using TI and BAR are superimposed in Figure S5. The $\Delta \Delta G$ trends are calculated for two target-E3 pairs, BTK-CRBN \textbf{(a)} and BRD4\textsuperscript{BD2}-VHL \textbf{(b)}. Solid lines represent the baseline $\Delta \Delta G$ trends where only volume exclusion is modeled between the two proteins, and dotted lines show the trends where nonspecific attractions are added. \textcolor{black}{The strengths ($\epsilon_\text{LJ}$) of  attractions are indicated by different colors. Higher $\epsilon_\text{LJ}$ represents stronger attractions,} and the baselines can also be considered as results at $\epsilon_\text{LJ}=0$. Results at $\epsilon_\text{LJ}=0.125$ and $0.2$ $kT$ are plotted for BTK-CRBN and results at $\epsilon_\text{LJ}=$0.125, 0.15, 0.175, 0.2, and 0.25 $kT$ are plotted for BRD4\textsuperscript{BD2}-VHL.} 
\label{fgr:BTK_CRBN}
\end{figure}

In the first setup, the steric cores of the proteins should penalize PROTAC binding and result in negative cooperativities. This is because some conformations that are accessible to the PROTAC in a binary PROTAC-protein complex become inaccessible in the ternary complex due to steric clashes. As the linker length increases and steric clashes are attenuated, the cooperativity should become less negative. We verify that such a monotonically increasing trend of negative $\Delta \Delta G$ is obtained in our model (Fig. \ref{fgr:BTK_CRBN}). 
Steric penalties on $\Delta \Delta G$ are most obvious at the region of short linker lengths (1-3 beads), after which the benefit from extending the linker length becomes increasingly marginal, and we expect that beyond the simulated window of linker lengths, $\Delta \Delta G$ will eventually plateau near 0. This $\Delta \Delta G$ trend is consistent with a recent effort to tabulate PROTAC linker length structure-activity relationships (SAR), which suggests that steric clashes at short linker lengths often result in a steep decrease in activity \cite{bemis_unraveling_2021}. 

After validating the baseline trend, we next examine how the cooperativity trend is changed by the addition of favorable PPIs through LJ potentials. \textcolor{black}{Increasing the well depth of LJ ($\epsilon_\text{LJ}$) increases the strength of this nonspecific attraction},
which is kept weak (Fig. S1) to approximate van der Waals forces. At the attraction strength of $\epsilon_\text{LJ}=0.125$ $kT$, the $\Delta \Delta G$ curve is elevated compared to the previous curve without attraction (Fig. \ref{fgr:BTK_CRBN}a), as favorable PPIs are expected to enhance cooperativity. Nevertheless, at this attraction strength, steric penalties still dominate and $\Delta \Delta G$s remains negative.
Even though adding an LJ potential brings an additional penalty when beads overlap, shorter PROTACs still benefit more from the attractive part of LJ than longer PROTACs, resulting in a flatter $\Delta \Delta G$ trend as compared with the purely repulsive interactions. 

An appropriate combination of repulsive and attractive forces may generate a non-monotonic $\Delta \Delta G$ trend, such that intermediate linker lengths promote optimal cooperativity by minimizing steric clashes while maximally sampling attractive PPIs\cite{bemis_unraveling_2021}. As the attraction strength increases to $\epsilon_\text{LJ}=0.2$ $kT$, intermediate-length PROTACs exhibit not only positive $\Delta \Delta G$s but the values can be comparable and even slightly higher than that of the longest PROTAC (Fig. \ref{fgr:BTK_CRBN}a). Within the limited window of linker lengths, only the initial part of the decaying tail of a non-monotonic $\Delta \Delta G$ trend is observed. We expect that beyond the simulated window of linker lengths, configurational entropic penalties will continue driving $\Delta \Delta G$ down towards 0. 

Experimentally, the linker length at 3 beads uniquely enables weak positive cooperativity for BTK-CRBN, whereas our results at $\epsilon_\text{LJ}=0.2$ $kT$ remain biased towards favoring longer linkers and are not as sensitive to linker length changes. To see whether these characteristics are specific to the choice of the system, we then examine the $\Delta \Delta G$ trends for a different system (Fig. \ref{fgr:BTK_CRBN}b), BRD4\textsuperscript{BD2}-VHL, where experimentally, the linker length at 3 beads can also optimize the cooperativity \cite{chan_impact_2018}. Due to the smaller size of the system, we can afford to calculate $\Delta \Delta G$s at three more attraction strengths. Similar to BTK-CRBN, in the absence of attractions, negative $\Delta \Delta G$ monotonically increases over the linker length, and 
adding nonspecific attractions results in flatter and higher $\Delta \Delta G$ curves. Within the narrow window of short linker lengths, scanning the attractive strength $\epsilon_\text{LJ}$ from 0.125 to 0.25 $kT$, however, does not recapitulate the optimal linker length at 3 beads. This result suggests that enhancing nonspecific attractions in the minimal model is insufficient to compensate for the steric penalties while remaining sensitive to entropic penalties from the linker length.

We demonstrate that the minimal CG model directly captures configurational entropic effects on weak nonspecific PPIs through analyzing $\Delta \Delta G$ trends over PROTAC linker lengths. Beyond this entropic effect, combining repulsive and attractive interactions at various strengths changes the behaviors of cooperativity trends and can shift the optimal linker length, as shown in BTK-CRBN. Nevertheless, chemically specific interactions or specific sampling of certain PPIs is needed to model optimal positive cooperativity at an experimentally relevant range and resolution of PROTAC linker lengths.

\subsection*{Electrostatics in PROTAC-mediated PPIs exhibit plasticity}

As a step towards more realistic modeling of cooperativity, we seek chemically specific PPIs to include and further explore the BRD4\textsuperscript{BD2}-VHL system due to the availability of experimental structural information. Crystal structures of the ternary complexes have revealed specific interactions that are proposed as the molecular basis for the observed positive cooperativity and \textcolor{black}{selectivity against other structural homologs.\cite{gadd_structural_2017,testa_structure-based_2020}}
As shown in the previous subsection, these interactions between proteins cannot be approximated by nonspecific attractions that contribute to the cooperativity with low sensitivity to linker length and no protein sequence dependence. 

\begin{figure}[htbp!]
\centering
\includegraphics[page=4,scale=0.48]{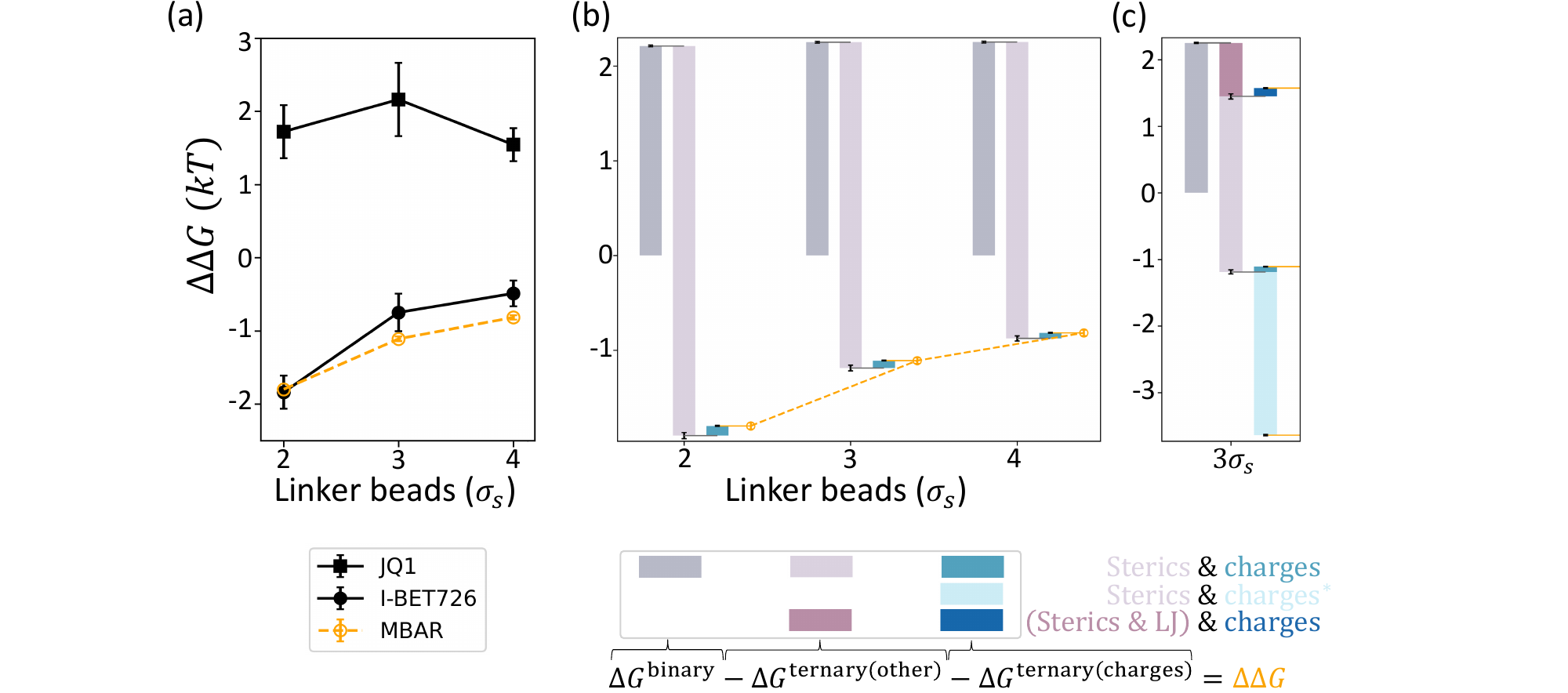}
\caption{ Electrostatic contributions to the cooperativity in the  BRD4\textsuperscript{BD2}-VHL system are small and context-dependent. All calculations shown are obtained using MBAR, and results using TI and BAR are shown in Figure S6.
\textbf{(a)} Calculations of $\Delta \Delta G$s over PROTAC linker lengths are shown with the experimental measurements \cite{chan_impact_2018} (black) converted to our units. Experimental results at 2, 3, and 4 linker beads correspond to MZ4, MZ1, and MZ2 for PROTACs using JQ1 warhead and MZP-61, MZP-54, MZP-55 for PROTACs using I-BET726 warhead. 
\textbf{(b)} Waterfall plot breakdown of $\Delta \Delta G$ calculations. At each linker length, bars in each triplet  correspond to $\Delta G\textsuperscript{binary}$ (grey), $- \Delta G\textsuperscript{ternary(other)}$ (light purple), and $-\Delta G\textsuperscript{ternary(charges)}$ (turquoise), and are arranged in a cumulative manner such that the end position marks the resulted $\Delta \Delta G$ (orange). $\Delta G\textsuperscript{ternary(other)}$ denotes the free energy change of turning on interaction energy components other than the electrostatics, which only include steric repulsions in this panel.
\textbf{(c)} $\Delta \Delta G$ breakdowns at linker length 3 under different forcefield parameterizations are superimposed for comparison. Reducing the screening effect by ten-fold (charges*) significantly increases $\Delta G\textsuperscript{ternary(charges)}$ (cyan), which leads to a very negative $\Delta \Delta G$. Introducing non-specific attractions ($\epsilon_\text{LJ}=0.2$ $kT$) not only reduces $\Delta G\textsuperscript{ternary(other)}$ (dark purple) but also doubles $\Delta G\textsuperscript{ternary(charges)}$ (steel blue), resulting in a positive $\Delta \Delta G$.}
\label{fgr:BRD4BD2_VHL}
\end{figure}

The structural findings such as salt bridges at the PPI interface and the mutational studies involving charged residues on BRD4\textsuperscript{BD2} and homologs\cite{gadd_structural_2017} motivate us to approach chemical specificity through modeling electrostatic interactions. As CGMD uses an implicit solvent, we choose the Debye-Hückel (DH) potential to describe electrostatics in consideration of screening effects under physiological conditions. Within the BRD4\textsuperscript{BD2}-VHL system, incorporating charges of protein beads results in a monotonic trend of negative $\Delta \Delta G$s with increasing linker length, (Fig. \ref{fgr:BRD4BD2_VHL}a) similar to the baseline obtained using steric repulsions only (Fig. \ref{fgr:BTK_CRBN}b). Since charges are perturbed separately in $\Delta \Delta G$ calculations for numeric stability, in the following discussions, we further investigate our $\Delta \Delta G$ results by isolating the final stage ($\Delta G\textsuperscript{ternary(charges)}$) in which charges are turned on in the presence of sterics.

Breaking down the $\Delta \Delta G$s by each \textcolor{black}{energy component} shows that at all three linker lengths, $\Delta G\textsuperscript{ternary(charges)}$ is slightly negative, indicating a mildly favorable process, but the penalty from steric repulsions overwhelmingly dominates electrostatic contributions by an order of magnitude (Fig. \ref{fgr:BRD4BD2_VHL}c). As PROTAC linker length increases from 2 to 4 beads, the contribution from $\Delta G\textsuperscript{ternary(charges)}$ monotonically diminishes. We consider the possibility that the screening of charges is too strong to model more favorable PPIs and tune the screening parameter in the DH potential at the linker length of 3 beads. However, because both the target protein and the E3 ligase have net positive charges, significantly weakening the screening strength leads to a much more unfavorable $\Delta G\textsuperscript{ternary(charges)}$ (Fig. \ref{fgr:BRD4BD2_VHL}c). 
It is also possible that our level of coarse-graining loses the spatial resolution required for this system to capture detailed interactions like salt bridge formation as observed in the crystal structures\cite{gadd_structural_2017,testa_structure-based_2020}. 

In addition to the small contribution to $\Delta \Delta G$, $\Delta G\textsuperscript{ternary(charges)}$ itself exhibits plasticity because conformational sampling at the stage of charge perturbation in alchemical free energy calculations is biased by the potentials turned on in previous stages. The presence of steric repulsions combined with nonspecific attractions at the strength of $\epsilon_\text{LJ}=0.2$ $kT$, for example, has doubled the $\Delta G\textsuperscript{ternary(charges)}$ obtained at the linker length of 3 beads without nonspecific attractions (Fig. \ref{fgr:BRD4BD2_VHL}c). 
Interestingly, this change in $\Delta G\textsuperscript{ternary(charges)}$ is on top of the favorable contribution from nonspecific attractions in the previous calculation stage ($\Delta G\textsuperscript{ternary(other)}$) before the inclusion of protein charges. For this particular ternary complex, nonspecific attractions and electrostatic interactions work synergistically.

Our dissection of the electrostatic component in $\Delta \Delta G$ under different simulation setups suggests that a more holistic parameterization is needed to accurately evaluate chemically specific PPIs. For BRD4\textsuperscript{BD2}-VHL, incorporating hydrophobic interactions will be of particular interest as there is stacking of hydrophobic residues at the PPI interface observed in the crystal structures.\cite{gadd_structural_2017,testa_structure-based_2020} Hydrophobic interactions may also introduce non-additive free-energy contributions with electrostatics in a similar manner seen with the nonspecific attractions.
It is also worth noting that the favorable PPIs revealed by crystal structures are enabled by PROTACs using a JQ1 warhead, which imposes a different linker attachment angle (i.e. exit vector) from an I-BET726 warhead (Fig. S7).\cite{chan_impact_2018}
Our current forcefield does not model the PROTAC linker with angular terms to specify the exit vectors, which  
leads to a $\Delta \Delta G$ trend that matches well with the worse-performing I-BET726 set of PROTACs (Fig. \ref{fgr:BRD4BD2_VHL}a). As rigidifying PROTACs is a common strategy to optimize the cooperativity by entropically enhancing certain PPIs\cite{testa_structure-based_2020,schiemer_snapshots_2021}, parameterizing linker conformations 
will improve modeling the specificity in PROTAC-mediated PPIs.

\section{Conclusions}  
We explore a novel computational approach to model the binding cooperativity of PROTACs by combining CGMD and alchemical free energy calculations. The plasticity of PROTAC-mediated PPIs motivates an unconventional application of alchemical methods at a perturbation scale that is rarely attempted. We show that with coarse-graining, converged estimates from various free energy calculation methods are attainable within a reasonable amount of computation time. Our results expand the possibility of more creative use of alchemical methods.
The feasibility and efficiency of the CG alchemical approach enable us to probe multiple energy components under the alchemical framework and characterize how PROTAC linker lengths modulate PPIs under different setups to produce distinct cooperativity trends. In addition to validating the benefit of using long linkers to avoid steric clashes, we demonstrate with a simple addition of nonspecific attractions between BTK and CRBN that the binding cooperativity can be promoted by shortening the PROTAC linker. 
Our minimal model is capable of unveiling such changes in cooperativity that are rooted in the configurational freedom of the ternary complexes rather than chemical properties. 

Quantitative modeling of the cooperativity, however, remains difficult due to the lack of specificity in the minimal model.
Previous studies have recognized the challenges brought by non-native PROTAC-mediated PPIs that are often weak, transient, and pliable, and have called for a paradigm shift towards an ensemble-based characterization beyond a handful of docked or crystal poses. \cite{nowak_plasticity_2018,bai_rationalizing_2021,eron_structural_2021}. 
While thermodynamic properties such as the binding cooperativity are inherently ensemble-based, we note that both accurate sampling of PPI conformations according to chemical properties and efficient computation to sample a diverse set of conformations are important for calculations. 
Currently, tuning the strength of nonspecific attractions cannot approximate favorable PPIs while retaining sensitivity to entropic constraints from the PROTAC linker length.
Simply adding electrostatic interactions based on amino acid charges proved insufficient to capture the cooperativity trend enabled by JQ1-based PROTACs in BRD4\textsuperscript{BD2}-VHL. 
Additional parameterizations are needed to capture chemically specific PPIs. 

Two main avenues are worth exploring for future improvement of our method -- PROTAC linker conformations and protein sequence-dependence. 
Among a myriad of PROTAC properties \cite{troup_current_2020} that we leave out, structural features such as the exit vector \cite{chan_impact_2018} and the linker rigidity \cite{testa_structure-based_2020,schiemer_snapshots_2021} in addition to the linker length can both entropically constrain the sampling of PPIs. 
Meanwhile, energy components of PPIs other than electrostatic interactions, notably the hydrophobic effects, are currently overlooked. Different energy components may have non-additive effects
in optimizing the absolute cooperativity and \textcolor{black}{relative cooperativities between target homologs such as BRD4\textsuperscript{BD2} and BRD4\textsuperscript{BD1}}.
Although coarse-graining enables efficient computation, parameterization for both directions of forcefield development will be a major hurdle to overcome. This can be bottom-up using shorter-timescale higher-resolution simulations, similar to that of the CG ENM (Fig. S2) in this work. 
A top-down fitting might also become possible with rapidly growing experimental studies that develop platforms \cite{hendrick_direct--biology_2022} for empirical SAR of PROTAC linkerology \cite{jmaple_developing_2019,ermondi_degraders_2020} or leverage promiscuous PROTACs and target homologs and mutants to investigate the molecular basis of specificity \cite{gopalsamy_selectivity_2022}.

\begin{suppinfo}

Description of the forcefield terms in CGMD, parameterization of CG ENM, analysis of phase space overlap in alchemical free energy calculations of the BTK-PROTAC (10)-CRBN complex, description of post-processing equilibrated and statistically de-correlated samples from CGMD trajectories for free energy calculations, convergence among TI, BAR, and MBAR for the results shown in Figure 3 and 4, and crystal structures showing the exit vector difference between JQ1 and I-BET726.

\end{suppinfo}

\begin{acknowledgement}

H.M. thanks William M. Clemons, Jr., Daniel Jacobson, Tomislav Begušić, Xuecheng Tao, Marta Gonzalvo, and Lixue Cheng for comments on the manuscript, and Zhen-Gang Wang and Christopher J. Balzer for technical discussions. We gratefully acknowledge support from the National Institutes of Health (NIH) R01GM138845 (8877\_CIT, subaward), Amgen Chem-Bio-Engineering Award (CBEA), and DeLogi Trust Science and Technology Grant. This work used the Extreme Science and Engineering Discovery Environment (XSEDE) Bridges computer at the Pittsburgh Supercomputing Center through allocation MCB160013 \cite{towns_xsede_2014}. XSEDE is supported by National Science Foundation grant number ACI-1548562. This work also used computational resources from the Resnick High Performance Computing Center, a facility supported by Resnick Sustainability Institute at the California Institute of Technology. 

\end{acknowledgement}

\bibliography{references}

\providecommand{\latin}[1]{#1}
\makeatletter
\providecommand{\doi}
  {\begingroup\let\do\@makeother\dospecials
  \catcode`\{=1 \catcode`\}=2 \doi@aux}
\providecommand{\doi@aux}[1]{\endgroup\texttt{#1}}
\makeatother
\providecommand*\mcitethebibliography{\thebibliography}
\csname @ifundefined\endcsname{endmcitethebibliography}
  {\let\endmcitethebibliography\endthebibliography}{}
\begin{mcitethebibliography}{68}
\providecommand*\natexlab[1]{#1}
\providecommand*\mciteSetBstSublistMode[1]{}
\providecommand*\mciteSetBstMaxWidthForm[2]{}
\providecommand*\mciteBstWouldAddEndPuncttrue
  {\def\EndOfBibitem{\unskip.}}
\providecommand*\mciteBstWouldAddEndPunctfalse
  {\let\EndOfBibitem\relax}
\providecommand*\mciteSetBstMidEndSepPunct[3]{}
\providecommand*\mciteSetBstSublistLabelBeginEnd[3]{}
\providecommand*\EndOfBibitem{}
\mciteSetBstSublistMode{f}
\mciteSetBstMaxWidthForm{subitem}{(\alph{mcitesubitemcount})}
\mciteSetBstSublistLabelBeginEnd
  {\mcitemaxwidthsubitemform\space}
  {\relax}
  {\relax}

\bibitem[An and Fu(2018)An, and Fu]{an_small-molecule_2018}
An,~S.; Fu,~L. Small-molecule {PROTACs}: {An} emerging and promising approach
  for the development of targeted therapy drugs. \emph{EBioMedicine}
  \textbf{2018}, \emph{36}, 553--562\relax
\mciteBstWouldAddEndPuncttrue
\mciteSetBstMidEndSepPunct{\mcitedefaultmidpunct}
{\mcitedefaultendpunct}{\mcitedefaultseppunct}\relax
\EndOfBibitem
\bibitem[Burslem and Crews(2020)Burslem, and
  Crews]{burslem_proteolysis-targeting_2020}
Burslem,~G.~M.; Crews,~C.~M. Proteolysis-{Targeting} {Chimeras} as
  {Therapeutics} and {Tools} for {Biological} {Discovery}. \emph{Cell}
  \textbf{2020}, \emph{181}, 102--114\relax
\mciteBstWouldAddEndPuncttrue
\mciteSetBstMidEndSepPunct{\mcitedefaultmidpunct}
{\mcitedefaultendpunct}{\mcitedefaultseppunct}\relax
\EndOfBibitem
\bibitem[Sakamoto \latin{et~al.}(2001)Sakamoto, Kim, Kumagai, Mercurio, Crews,
  and Deshaies]{sakamoto_protacs_2001}
Sakamoto,~K.~M.; Kim,~K.~B.; Kumagai,~A.; Mercurio,~F.; Crews,~C.~M.;
  Deshaies,~R.~J. Protacs: {Chimeric} molecules that target proteins to the
  {Skp1}–{Cullin}–{F} box complex for ubiquitination and degradation.
  \emph{Proceedings of the National Academy of Sciences} \textbf{2001},
  \emph{98}, 8554--8559, Publisher: National Academy of Sciences Section:
  Biological Sciences\relax
\mciteBstWouldAddEndPuncttrue
\mciteSetBstMidEndSepPunct{\mcitedefaultmidpunct}
{\mcitedefaultendpunct}{\mcitedefaultseppunct}\relax
\EndOfBibitem
\bibitem[Mullard(2021)]{mullard_targeted_2021}
Mullard,~A. Targeted protein degraders crowd into the clinic. \emph{Nature
  Reviews Drug Discovery} \textbf{2021}, \emph{20}, 247--250, Bandiera\_abtest:
  a Cg\_type: News Number: 4 Publisher: Nature Publishing Group\relax
\mciteBstWouldAddEndPuncttrue
\mciteSetBstMidEndSepPunct{\mcitedefaultmidpunct}
{\mcitedefaultendpunct}{\mcitedefaultseppunct}\relax
\EndOfBibitem
\bibitem[Maniaci and Ciulli(2019)Maniaci, and
  Ciulli]{maniaci_bifunctional_2019}
Maniaci,~C.; Ciulli,~A. Bifunctional chemical probes inducing protein–protein
  interactions. \emph{Current Opinion in Chemical Biology} \textbf{2019},
  \emph{52}, 145--156\relax
\mciteBstWouldAddEndPuncttrue
\mciteSetBstMidEndSepPunct{\mcitedefaultmidpunct}
{\mcitedefaultendpunct}{\mcitedefaultseppunct}\relax
\EndOfBibitem
\bibitem[Troup \latin{et~al.}(2020)Troup, Fallan, and Baud]{troup_current_2020}
Troup,~R.~I.; Fallan,~C.; Baud,~M. G.~J. Current strategies for the design of
  {PROTAC} linkers: a critical review. \emph{Exploration of Targeted Anti-tumor
  Therapy} \textbf{2020}, \emph{1}, 273--312, Publisher: Open Exploration\relax
\mciteBstWouldAddEndPuncttrue
\mciteSetBstMidEndSepPunct{\mcitedefaultmidpunct}
{\mcitedefaultendpunct}{\mcitedefaultseppunct}\relax
\EndOfBibitem
\bibitem[Alabi and Crews(2021)Alabi, and Crews]{alabi_major_2021}
Alabi,~S.; Crews,~C. Major {Advances} in {Targeted} {Protein} {Degradation}:
  {PROTACs}, {LYTACs}, and {MADTACs}. \emph{Journal of Biological Chemistry}
  \textbf{2021}, 100647\relax
\mciteBstWouldAddEndPuncttrue
\mciteSetBstMidEndSepPunct{\mcitedefaultmidpunct}
{\mcitedefaultendpunct}{\mcitedefaultseppunct}\relax
\EndOfBibitem
\bibitem[Gadd \latin{et~al.}(2017)Gadd, Testa, Lucas, Chan, Chen, Lamont,
  Zengerle, and Ciulli]{gadd_structural_2017}
Gadd,~M.~S.; Testa,~A.; Lucas,~X.; Chan,~K.-H.; Chen,~W.; Lamont,~D.~J.;
  Zengerle,~M.; Ciulli,~A. Structural basis of {PROTAC} cooperative recognition
  for selective protein degradation. \emph{Nature Chemical Biology}
  \textbf{2017}, \emph{13}, 514--521, Number: 5 Publisher: Nature Publishing
  Group\relax
\mciteBstWouldAddEndPuncttrue
\mciteSetBstMidEndSepPunct{\mcitedefaultmidpunct}
{\mcitedefaultendpunct}{\mcitedefaultseppunct}\relax
\EndOfBibitem
\bibitem[Nowak \latin{et~al.}(2018)Nowak, DeAngelo, Buckley, He, Donovan, An,
  Safaee, Jedrychowski, Ponthier, Ishoey, Zhang, Mancias, Gray, Bradner, and
  Fischer]{nowak_plasticity_2018}
Nowak,~R.~P.; DeAngelo,~S.~L.; Buckley,~D.; He,~Z.; Donovan,~K.~A.; An,~J.;
  Safaee,~N.; Jedrychowski,~M.~P.; Ponthier,~C.~M.; Ishoey,~M. \latin{et~al.}
  Plasticity in binding confers selectivity in ligand-induced protein
  degradation. \emph{Nature Chemical Biology} \textbf{2018}, \emph{14},
  706--714, Number: 7 Publisher: Nature Publishing Group\relax
\mciteBstWouldAddEndPuncttrue
\mciteSetBstMidEndSepPunct{\mcitedefaultmidpunct}
{\mcitedefaultendpunct}{\mcitedefaultseppunct}\relax
\EndOfBibitem
\bibitem[Smith \latin{et~al.}(2019)Smith, Wang, Jaime-Figueroa, Harbin, Wang,
  Hamman, and Crews]{smith_differential_2019}
Smith,~B.~E.; Wang,~S.~L.; Jaime-Figueroa,~S.; Harbin,~A.; Wang,~J.;
  Hamman,~B.~D.; Crews,~C.~M. Differential {PROTAC} substrate specificity
  dictated by orientation of recruited {E3} ligase. \emph{Nature
  Communications} \textbf{2019}, \emph{10}, 131, Number: 1 Publisher: Nature
  Publishing Group\relax
\mciteBstWouldAddEndPuncttrue
\mciteSetBstMidEndSepPunct{\mcitedefaultmidpunct}
{\mcitedefaultendpunct}{\mcitedefaultseppunct}\relax
\EndOfBibitem
\bibitem[Riching \latin{et~al.}(2018)Riching, Mahan, Corona, McDougall, Vasta,
  Robers, Urh, and Daniels]{riching_quantitative_2018}
Riching,~K.~M.; Mahan,~S.; Corona,~C.~R.; McDougall,~M.; Vasta,~J.~D.;
  Robers,~M.~B.; Urh,~M.; Daniels,~D.~L. Quantitative {Live}-{Cell} {Kinetic}
  {Degradation} and {Mechanistic} {Profiling} of {PROTAC} {Mode} of {Action}.
  \emph{ACS Chemical Biology} \textbf{2018}, \emph{13}, 2758--2770, Publisher:
  American Chemical Society\relax
\mciteBstWouldAddEndPuncttrue
\mciteSetBstMidEndSepPunct{\mcitedefaultmidpunct}
{\mcitedefaultendpunct}{\mcitedefaultseppunct}\relax
\EndOfBibitem
\bibitem[Roy \latin{et~al.}(2019)Roy, Winkler, Hughes, Whitworth, Galant,
  Farnaby, Rumpel, and Ciulli]{roy_spr-measured_2019}
Roy,~M.~J.; Winkler,~S.; Hughes,~S.~J.; Whitworth,~C.; Galant,~M.; Farnaby,~W.;
  Rumpel,~K.; Ciulli,~A. {SPR}-{Measured} {Dissociation} {Kinetics} of {PROTAC}
  {Ternary} {Complexes} {Influence} {Target} {Degradation} {Rate}. \emph{ACS
  Chemical Biology} \textbf{2019}, \emph{14}, 361--368, Publisher: American
  Chemical Society\relax
\mciteBstWouldAddEndPuncttrue
\mciteSetBstMidEndSepPunct{\mcitedefaultmidpunct}
{\mcitedefaultendpunct}{\mcitedefaultseppunct}\relax
\EndOfBibitem
\bibitem[Farnaby \latin{et~al.}(2019)Farnaby, Koegl, Roy, Whitworth, Diers,
  Trainor, Zollman, Steurer, Karolyi-Oezguer, Riedmueller, Gmaschitz, Wachter,
  Dank, Galant, Sharps, Rumpel, Traxler, Gerstberger, Schnitzer, Petermann,
  Greb, Weinstabl, Bader, Zoephel, Weiss-Puxbaum, Ehrenhöfer-Wölfer, Wöhrle,
  Boehmelt, Rinnenthal, Arnhof, Wiechens, Wu, Owen-Hughes, Ettmayer, Pearson,
  McConnell, and Ciulli]{farnaby_baf_2019}
Farnaby,~W.; Koegl,~M.; Roy,~M.~J.; Whitworth,~C.; Diers,~E.; Trainor,~N.;
  Zollman,~D.; Steurer,~S.; Karolyi-Oezguer,~J.; Riedmueller,~C. \latin{et~al.}
   {BAF} complex vulnerabilities in cancer demonstrated via structure-based
  {PROTAC} design. \emph{Nature Chemical Biology} \textbf{2019}, \emph{15},
  672--680, Number: 7 Publisher: Nature Publishing Group\relax
\mciteBstWouldAddEndPuncttrue
\mciteSetBstMidEndSepPunct{\mcitedefaultmidpunct}
{\mcitedefaultendpunct}{\mcitedefaultseppunct}\relax
\EndOfBibitem
\bibitem[Du \latin{et~al.}(2019)Du, Volkov, Czerwinski, Tan, Huerta, Morton,
  Rizzi, Wehn, Xu, Nijhawan, and Wallace]{du_structural_2019}
Du,~X.; Volkov,~O.~A.; Czerwinski,~R.~M.; Tan,~H.; Huerta,~C.; Morton,~E.~R.;
  Rizzi,~J.~P.; Wehn,~P.~M.; Xu,~R.; Nijhawan,~D. \latin{et~al.}  Structural
  {Basis} and {Kinetic} {Pathway} of {RBM39} {Recruitment} to {DCAF15} by a
  {Sulfonamide} {Molecular} {Glue} {E7820}. \emph{Structure} \textbf{2019},
  \emph{27}, 1625--1633.e3\relax
\mciteBstWouldAddEndPuncttrue
\mciteSetBstMidEndSepPunct{\mcitedefaultmidpunct}
{\mcitedefaultendpunct}{\mcitedefaultseppunct}\relax
\EndOfBibitem
\bibitem[Lai \latin{et~al.}(2016)Lai, Toure, Hellerschmied, Salami,
  Jaime-Figueroa, Ko, Hines, and Crews]{lai_modular_2016}
Lai,~A.~C.; Toure,~M.; Hellerschmied,~D.; Salami,~J.; Jaime-Figueroa,~S.;
  Ko,~E.; Hines,~J.; Crews,~C.~M. Modular {PROTAC} {Design} for the
  {Degradation} of {Oncogenic} {BCR}-{ABL}. \emph{Angewandte Chemie
  International Edition} \textbf{2016}, \emph{55}, 807--810, \_eprint:
  https://onlinelibrary.wiley.com/doi/pdf/10.1002/anie.201507634\relax
\mciteBstWouldAddEndPuncttrue
\mciteSetBstMidEndSepPunct{\mcitedefaultmidpunct}
{\mcitedefaultendpunct}{\mcitedefaultseppunct}\relax
\EndOfBibitem
\bibitem[Bondeson \latin{et~al.}(2018)Bondeson, Smith, Burslem, Buhimschi,
  Hines, Jaime-Figueroa, Wang, Hamman, Ishchenko, and
  Crews]{bondeson_lessons_2018}
Bondeson,~D.~P.; Smith,~B.~E.; Burslem,~G.~M.; Buhimschi,~A.~D.; Hines,~J.;
  Jaime-Figueroa,~S.; Wang,~J.; Hamman,~B.~D.; Ishchenko,~A.; Crews,~C.~M.
  Lessons in {PROTAC} {Design} from {Selective} {Degradation} with a
  {Promiscuous} {Warhead}. \emph{Cell Chemical Biology} \textbf{2018},
  \emph{25}, 78--87.e5\relax
\mciteBstWouldAddEndPuncttrue
\mciteSetBstMidEndSepPunct{\mcitedefaultmidpunct}
{\mcitedefaultendpunct}{\mcitedefaultseppunct}\relax
\EndOfBibitem
\bibitem[Donovan \latin{et~al.}(2020)Donovan, Ferguson, Bushman, Eleuteri,
  Bhunia, Ryu, Tan, Shi, Yue, Liu, Dobrovolsky, Jiang, Wang, Hao, You, Teng,
  Liang, Hatcher, Li, Manz, Groendyke, Hu, Nam, Sengupta, Cho, Shin, Agius,
  Ghobrial, Ma, Che, Buhrlage, Sim, Gray, and Fischer]{donovan_mapping_2020}
Donovan,~K.~A.; Ferguson,~F.~M.; Bushman,~J.~W.; Eleuteri,~N.~A.; Bhunia,~D.;
  Ryu,~S.; Tan,~L.; Shi,~K.; Yue,~H.; Liu,~X. \latin{et~al.}  Mapping the
  {Degradable} {Kinome} {Provides} a {Resource} for {Expedited} {Degrader}
  {Development}. \emph{Cell} \textbf{2020}, \emph{183}, 1714--1731.e10\relax
\mciteBstWouldAddEndPuncttrue
\mciteSetBstMidEndSepPunct{\mcitedefaultmidpunct}
{\mcitedefaultendpunct}{\mcitedefaultseppunct}\relax
\EndOfBibitem
\bibitem[Spradlin \latin{et~al.}(2019)Spradlin, Hu, Ward, Brittain, Jones, Ou,
  To, Proudfoot, Ornelas, Woldegiorgis, Olzmann, Bussiere, Thomas, Tallarico,
  McKenna, Schirle, Maimone, and Nomura]{spradlin_harnessing_2019}
Spradlin,~J.~N.; Hu,~X.; Ward,~C.~C.; Brittain,~S.~M.; Jones,~M.~D.; Ou,~L.;
  To,~M.; Proudfoot,~A.; Ornelas,~E.; Woldegiorgis,~M. \latin{et~al.}
  Harnessing the anti-cancer natural product nimbolide for targeted protein
  degradation. \emph{Nature Chemical Biology} \textbf{2019}, \emph{15},
  747--755, Number: 7 Publisher: Nature Publishing Group\relax
\mciteBstWouldAddEndPuncttrue
\mciteSetBstMidEndSepPunct{\mcitedefaultmidpunct}
{\mcitedefaultendpunct}{\mcitedefaultseppunct}\relax
\EndOfBibitem
\bibitem[Ward \latin{et~al.}(2019)Ward, Kleinman, Brittain, Lee, Chung, Kim,
  Petri, Thomas, Tallarico, McKenna, Schirle, and Nomura]{ward_covalent_2019}
Ward,~C.~C.; Kleinman,~J.~I.; Brittain,~S.~M.; Lee,~P.~S.; Chung,~C. Y.~S.;
  Kim,~K.; Petri,~Y.; Thomas,~J.~R.; Tallarico,~J.~A.; McKenna,~J.~M.
  \latin{et~al.}  Covalent {Ligand} {Screening} {Uncovers} a {RNF4} {E3}
  {Ligase} {Recruiter} for {Targeted} {Protein} {Degradation} {Applications}.
  \emph{ACS Chemical Biology} \textbf{2019}, \emph{14}, 2430--2440, Publisher:
  American Chemical Society\relax
\mciteBstWouldAddEndPuncttrue
\mciteSetBstMidEndSepPunct{\mcitedefaultmidpunct}
{\mcitedefaultendpunct}{\mcitedefaultseppunct}\relax
\EndOfBibitem
\bibitem[Zhang \latin{et~al.}(2019)Zhang, Crowley, Wucherpfennig, Dix, and
  Cravatt]{zhang_electrophilic_2019}
Zhang,~X.; Crowley,~V.~M.; Wucherpfennig,~T.~G.; Dix,~M.~M.; Cravatt,~B.~F.
  Electrophilic {PROTACs} that degrade nuclear proteins by engaging {DCAF16}.
  \emph{Nature Chemical Biology} \textbf{2019}, \emph{15}, 737--746, Number: 7
  Publisher: Nature Publishing Group\relax
\mciteBstWouldAddEndPuncttrue
\mciteSetBstMidEndSepPunct{\mcitedefaultmidpunct}
{\mcitedefaultendpunct}{\mcitedefaultseppunct}\relax
\EndOfBibitem
\bibitem[Kuljanin \latin{et~al.}(2021)Kuljanin, Mitchell, Schweppe, Gikandi,
  Nusinow, Bulloch, Vinogradova, Wilson, Kool, Mancias, Cravatt, and
  Gygi]{kuljanin_reimagining_2021}
Kuljanin,~M.; Mitchell,~D.~C.; Schweppe,~D.~K.; Gikandi,~A.~S.; Nusinow,~D.~P.;
  Bulloch,~N.~J.; Vinogradova,~E.~V.; Wilson,~D.~L.; Kool,~E.~T.;
  Mancias,~J.~D. \latin{et~al.}  Reimagining high-throughput profiling of
  reactive cysteines for cell-based screening of large electrophile libraries.
  \emph{Nature Biotechnology} \textbf{2021}, \emph{39}, 630--641, Number: 5
  Publisher: Nature Publishing Group\relax
\mciteBstWouldAddEndPuncttrue
\mciteSetBstMidEndSepPunct{\mcitedefaultmidpunct}
{\mcitedefaultendpunct}{\mcitedefaultseppunct}\relax
\EndOfBibitem
\bibitem[Li \latin{et~al.}(2008)Li, Bengtson, Ulbrich, Matsuda, Reddy, Orth,
  Chanda, Batalov, and Joazeiro]{li_genome-wide_2008}
Li,~W.; Bengtson,~M.~H.; Ulbrich,~A.; Matsuda,~A.; Reddy,~V.~A.; Orth,~A.;
  Chanda,~S.~K.; Batalov,~S.; Joazeiro,~C. A.~P. Genome-{Wide} and {Functional}
  {Annotation} of {Human} {E3} {Ubiquitin} {Ligases} {Identifies} {MULAN}, a
  {Mitochondrial} {E3} that {Regulates} the {Organelle}'s {Dynamics} and
  {Signaling}. \emph{PLOS ONE} \textbf{2008}, \emph{3}, e1487, Publisher:
  Public Library of Science\relax
\mciteBstWouldAddEndPuncttrue
\mciteSetBstMidEndSepPunct{\mcitedefaultmidpunct}
{\mcitedefaultendpunct}{\mcitedefaultseppunct}\relax
\EndOfBibitem
\bibitem[Jevtić \latin{et~al.}(2021)Jevtić, Haakonsen, and
  Rapé]{jevtic_e3_2021}
Jevtić,~P.; Haakonsen,~D.~L.; Rapé,~M. An {E3} ligase guide to the galaxy of
  small-molecule-induced protein degradation. \emph{Cell Chemical Biology}
  \textbf{2021}, \relax
\mciteBstWouldAddEndPunctfalse
\mciteSetBstMidEndSepPunct{\mcitedefaultmidpunct}
{}{\mcitedefaultseppunct}\relax
\EndOfBibitem
\bibitem[Scholes \latin{et~al.}(2021)Scholes, Mayor-Ruiz, and
  Winter]{scholes_identification_2021}
Scholes,~N.~S.; Mayor-Ruiz,~C.; Winter,~G.~E. Identification and selectivity
  profiling of small-molecule degraders via multi-omics approaches. \emph{Cell
  Chemical Biology} \textbf{2021}, \relax
\mciteBstWouldAddEndPunctfalse
\mciteSetBstMidEndSepPunct{\mcitedefaultmidpunct}
{}{\mcitedefaultseppunct}\relax
\EndOfBibitem
\bibitem[Huang \latin{et~al.}(2018)Huang, Dobrovolsky, Paulk, Yang, Weisberg,
  Doctor, Buckley, Cho, Ko, Jang, Shi, Choi, Griffin, Li, Treon, Fischer,
  Bradner, Tan, and Gray]{huang_chemoproteomic_2018}
Huang,~H.-T.; Dobrovolsky,~D.; Paulk,~J.; Yang,~G.; Weisberg,~E.~L.;
  Doctor,~Z.~M.; Buckley,~D.~L.; Cho,~J.-H.; Ko,~E.; Jang,~J. \latin{et~al.}  A
  {Chemoproteomic} {Approach} to {Query} the {Degradable} {Kinome} {Using} a
  {Multi}-kinase {Degrader}. \emph{Cell Chemical Biology} \textbf{2018},
  \emph{25}, 88--99.e6\relax
\mciteBstWouldAddEndPuncttrue
\mciteSetBstMidEndSepPunct{\mcitedefaultmidpunct}
{\mcitedefaultendpunct}{\mcitedefaultseppunct}\relax
\EndOfBibitem
\bibitem[Rodriguez-Rivera and Levi(2021)Rodriguez-Rivera, and
  Levi]{rodriguez-rivera_unifying_2021}
Rodriguez-Rivera,~F.~P.; Levi,~S.~M. Unifying {Catalysis} {Framework} to
  {Dissect} {Proteasomal} {Degradation} {Paradigms}. \emph{ACS Central Science}
  \textbf{2021}, Publisher: American Chemical Society\relax
\mciteBstWouldAddEndPuncttrue
\mciteSetBstMidEndSepPunct{\mcitedefaultmidpunct}
{\mcitedefaultendpunct}{\mcitedefaultseppunct}\relax
\EndOfBibitem
\bibitem[Maniaci \latin{et~al.}(2017)Maniaci, Hughes, Testa, Chen, Lamont,
  Rocha, Alessi, Romeo, and Ciulli]{maniaci_homo-protacs_2017}
Maniaci,~C.; Hughes,~S.~J.; Testa,~A.; Chen,~W.; Lamont,~D.~J.; Rocha,~S.;
  Alessi,~D.~R.; Romeo,~R.; Ciulli,~A. Homo-{PROTACs}: bivalent small-molecule
  dimerizers of the {VHL} {E3} ubiquitin ligase to induce self-degradation.
  \emph{Nature Communications} \textbf{2017}, \emph{8}, 830, Number: 1
  Publisher: Nature Publishing Group\relax
\mciteBstWouldAddEndPuncttrue
\mciteSetBstMidEndSepPunct{\mcitedefaultmidpunct}
{\mcitedefaultendpunct}{\mcitedefaultseppunct}\relax
\EndOfBibitem
\bibitem[Chan \latin{et~al.}(2018)Chan, Zengerle, Testa, and
  Ciulli]{chan_impact_2018}
Chan,~K.-H.; Zengerle,~M.; Testa,~A.; Ciulli,~A. Impact of {Target} {Warhead}
  and {Linkage} {Vector} on {Inducing} {Protein} {Degradation}: {Comparison} of
  {Bromodomain} and {Extra}-{Terminal} ({BET}) {Degraders} {Derived} from
  {Triazolodiazepine} ({JQ1}) and {Tetrahydroquinoline} ({I}-{BET726}) {BET}
  {Inhibitor} {Scaffolds}. \emph{Journal of Medicinal Chemistry} \textbf{2018},
  \emph{61}, 504--513, Publisher: American Chemical Society\relax
\mciteBstWouldAddEndPuncttrue
\mciteSetBstMidEndSepPunct{\mcitedefaultmidpunct}
{\mcitedefaultendpunct}{\mcitedefaultseppunct}\relax
\EndOfBibitem
\bibitem[Zorba \latin{et~al.}(2018)Zorba, Nguyen, Xu, Starr, Borzilleri, Smith,
  Zhu, Farley, Ding, Schiemer, Feng, Chang, Uccello, Young, Garcia-Irrizary,
  Czabaniuk, Schuff, Oliver, Montgomery, Hayward, Coe, Chen, Niosi, Luthra,
  Shah, El-Kattan, Qiu, West, Noe, Shanmugasundaram, Gilbert, Brown, and
  Calabrese]{zorba_delineating_2018}
Zorba,~A.; Nguyen,~C.; Xu,~Y.; Starr,~J.; Borzilleri,~K.; Smith,~J.; Zhu,~H.;
  Farley,~K.~A.; Ding,~W.; Schiemer,~J. \latin{et~al.}  Delineating the role of
  cooperativity in the design of potent {PROTACs} for {BTK}. \emph{Proceedings
  of the National Academy of Sciences} \textbf{2018}, \emph{115}, E7285--E7292,
  ISBN: 9781803662114 Publisher: National Academy of Sciences Section: PNAS
  Plus\relax
\mciteBstWouldAddEndPuncttrue
\mciteSetBstMidEndSepPunct{\mcitedefaultmidpunct}
{\mcitedefaultendpunct}{\mcitedefaultseppunct}\relax
\EndOfBibitem
\bibitem[Schiemer \latin{et~al.}(2021)Schiemer, Horst, Meng, Montgomery, Xu,
  Feng, Borzilleri, Uccello, Leverett, Brown, Che, Brown, Hayward, Gilbert,
  Noe, and Calabrese]{schiemer_snapshots_2021}
Schiemer,~J.; Horst,~R.; Meng,~Y.; Montgomery,~J.~I.; Xu,~Y.; Feng,~X.;
  Borzilleri,~K.; Uccello,~D.~P.; Leverett,~C.; Brown,~S. \latin{et~al.}
  Snapshots and ensembles of {BTK} and {cIAP1} protein degrader ternary
  complexes. \emph{Nature Chemical Biology} \textbf{2021}, \emph{17}, 152--160,
  Number: 2 Publisher: Nature Publishing Group\relax
\mciteBstWouldAddEndPuncttrue
\mciteSetBstMidEndSepPunct{\mcitedefaultmidpunct}
{\mcitedefaultendpunct}{\mcitedefaultseppunct}\relax
\EndOfBibitem
\bibitem[Drummond \latin{et~al.}(2020)Drummond, Henry, Li, and
  Williams]{drummond_improved_2020}
Drummond,~M.~L.; Henry,~A.; Li,~H.; Williams,~C.~I. Improved {Accuracy} for
  {Modeling} {PROTAC}-{Mediated} {Ternary} {Complex} {Formation} and {Targeted}
  {Protein} {Degradation} via {New} {In} {Silico} {Methodologies}.
  \emph{Journal of Chemical Information and Modeling} \textbf{2020}, \emph{60},
  5234--5254, Publisher: American Chemical Society\relax
\mciteBstWouldAddEndPuncttrue
\mciteSetBstMidEndSepPunct{\mcitedefaultmidpunct}
{\mcitedefaultendpunct}{\mcitedefaultseppunct}\relax
\EndOfBibitem
\bibitem[Zaidman \latin{et~al.}(2020)Zaidman, Prilusky, and
  London]{zaidman_prosettac_2020}
Zaidman,~D.; Prilusky,~J.; London,~N. {PRosettaC}: {Rosetta} {Based} {Modeling}
  of {PROTAC} {Mediated} {Ternary} {Complexes}. \emph{Journal of Chemical
  Information and Modeling} \textbf{2020}, \emph{60}, 4894--4903, Publisher:
  American Chemical Society\relax
\mciteBstWouldAddEndPuncttrue
\mciteSetBstMidEndSepPunct{\mcitedefaultmidpunct}
{\mcitedefaultendpunct}{\mcitedefaultseppunct}\relax
\EndOfBibitem
\bibitem[Weng \latin{et~al.}(2021)Weng, Li, Kang, and
  Hou]{weng_integrative_2021}
Weng,~G.; Li,~D.; Kang,~Y.; Hou,~T. Integrative {Modeling} of
  {PROTAC}-{Mediated} {Ternary} {Complexes}. \emph{Journal of Medicinal
  Chemistry} \textbf{2021}, \emph{64}, 16271--16281, Publisher: American
  Chemical Society\relax
\mciteBstWouldAddEndPuncttrue
\mciteSetBstMidEndSepPunct{\mcitedefaultmidpunct}
{\mcitedefaultendpunct}{\mcitedefaultseppunct}\relax
\EndOfBibitem
\bibitem[Bai \latin{et~al.}(2021)Bai, Miller, Andrianov, Yates, Kirubakaran,
  and Karanicolas]{bai_rationalizing_2021}
Bai,~N.; Miller,~S.~A.; Andrianov,~G.~V.; Yates,~M.; Kirubakaran,~P.;
  Karanicolas,~J. Rationalizing {PROTAC}-{Mediated} {Ternary} {Complex}
  {Formation} {Using} {Rosetta}. \emph{Journal of Chemical Information and
  Modeling} \textbf{2021}, \emph{61}, 1368--1382, Publisher: American Chemical
  Society\relax
\mciteBstWouldAddEndPuncttrue
\mciteSetBstMidEndSepPunct{\mcitedefaultmidpunct}
{\mcitedefaultendpunct}{\mcitedefaultseppunct}\relax
\EndOfBibitem
\bibitem[Bai \latin{et~al.}(2022)Bai, Riching, Makaju, Wu, Acker, Ou, Zhang,
  Shen, Bulloch, Rui, Gibson, Daniels, Urh, Rock, and
  Humphreys]{bai_modeling_2022}
Bai,~N.; Riching,~K.~M.; Makaju,~A.; Wu,~H.; Acker,~T.~M.; Ou,~S.-C.;
  Zhang,~Y.; Shen,~X.; Bulloch,~D.; Rui,~H. \latin{et~al.}  Modeling the
  {CRL4A} ligase complex to predict target protein ubiquitination induced by
  cereblon-recruiting {PROTACs}. \emph{Journal of Biological Chemistry}
  \textbf{2022}, 101653\relax
\mciteBstWouldAddEndPuncttrue
\mciteSetBstMidEndSepPunct{\mcitedefaultmidpunct}
{\mcitedefaultendpunct}{\mcitedefaultseppunct}\relax
\EndOfBibitem
\bibitem[Moreira \latin{et~al.}(2010)Moreira, Fernandes, and
  Ramos]{moreira_proteinprotein_2010}
Moreira,~I.~S.; Fernandes,~P.~A.; Ramos,~M.~J. Protein–protein docking
  dealing with the unknown. \emph{Journal of Computational Chemistry}
  \textbf{2010}, \emph{31}, 317--342, \_eprint:
  https://onlinelibrary.wiley.com/doi/pdf/10.1002/jcc.21276\relax
\mciteBstWouldAddEndPuncttrue
\mciteSetBstMidEndSepPunct{\mcitedefaultmidpunct}
{\mcitedefaultendpunct}{\mcitedefaultseppunct}\relax
\EndOfBibitem
\bibitem[Gromiha \latin{et~al.}(2017)Gromiha, Yugandhar, and
  Jemimah]{gromiha_proteinprotein_2017}
Gromiha,~M.~M.; Yugandhar,~K.; Jemimah,~S. Protein–protein interactions:
  scoring schemes and binding affinity. \emph{Current Opinion in Structural
  Biology} \textbf{2017}, \emph{44}, 31--38\relax
\mciteBstWouldAddEndPuncttrue
\mciteSetBstMidEndSepPunct{\mcitedefaultmidpunct}
{\mcitedefaultendpunct}{\mcitedefaultseppunct}\relax
\EndOfBibitem
\bibitem[Bemis \latin{et~al.}(2021)Bemis, La~Clair, and
  Burkart]{bemis_unraveling_2021}
Bemis,~T.~A.; La~Clair,~J.~J.; Burkart,~M.~D. Unraveling the {Role} of {Linker}
  {Design} in {Proteolysis} {Targeting} {Chimeras}. \emph{Journal of Medicinal
  Chemistry} \textbf{2021}, Publisher: American Chemical Society\relax
\mciteBstWouldAddEndPuncttrue
\mciteSetBstMidEndSepPunct{\mcitedefaultmidpunct}
{\mcitedefaultendpunct}{\mcitedefaultseppunct}\relax
\EndOfBibitem
\bibitem[Eron \latin{et~al.}(2021)Eron, Huang, Agafonov, Fitzgerald, Patel,
  Michael, Lee, Hart, Shaulsky, Nasveschuk, Phillips, Fisher, and
  Good]{eron_structural_2021}
Eron,~S.~J.; Huang,~H.; Agafonov,~R.~V.; Fitzgerald,~M.~E.; Patel,~J.;
  Michael,~R.~E.; Lee,~T.~D.; Hart,~A.~A.; Shaulsky,~J.; Nasveschuk,~C.~G.
  \latin{et~al.}  Structural {Characterization} of {Degrader}-{Induced}
  {Ternary} {Complexes} {Using} {Hydrogen}–{Deuterium} {Exchange} {Mass}
  {Spectrometry} and {Computational} {Modeling}: {Implications} for
  {Structure}-{Based} {Design}. \emph{ACS Chemical Biology} \textbf{2021},
  Publisher: American Chemical Society\relax
\mciteBstWouldAddEndPuncttrue
\mciteSetBstMidEndSepPunct{\mcitedefaultmidpunct}
{\mcitedefaultendpunct}{\mcitedefaultseppunct}\relax
\EndOfBibitem
\bibitem[Dixon \latin{et~al.}(2021)Dixon, MacPherson, Mostofian, Dauzhenka,
  Lotz, McGee, Shechter, Shrestha, Wiewiora, McDargh, Pei, Pal, Ribeiro,
  Wilkerson, Sachdeva, Gao, Jain, Sparks, Li, Vinitsky, Razavi, Kolossváry,
  Imbriglio, Evdokimov, Bergeron, Dickson, Xu, Sherman, and
  Izaguirre]{dixon_atomic-resolution_2021}
Dixon,~T.; MacPherson,~D.; Mostofian,~B.; Dauzhenka,~T.; Lotz,~S.; McGee,~D.;
  Shechter,~S.; Shrestha,~U.~R.; Wiewiora,~R.; McDargh,~Z.~A. \latin{et~al.}
  \emph{Atomic-{Resolution} {Prediction} of {Degrader}-mediated {Ternary}
  {Complex} {Structures} by {Combining} {Molecular} {Simulations} with
  {Hydrogen} {Deuterium} {Exchange}}; 2021; p 2021.09.26.461830, Section: New
  Results Type: article\relax
\mciteBstWouldAddEndPuncttrue
\mciteSetBstMidEndSepPunct{\mcitedefaultmidpunct}
{\mcitedefaultendpunct}{\mcitedefaultseppunct}\relax
\EndOfBibitem
\bibitem[Li \latin{et~al.}(2022)Li, Zhang, Guo, and Wang]{li_importance_2022}
Li,~W.; Zhang,~J.; Guo,~L.; Wang,~Q. Importance of {Three}-{Body} {Problems}
  and {Protein}–{Protein} {Interactions} in {Proteolysis}-{Targeting}
  {Chimera} {Modeling}: {Insights} from {Molecular} {Dynamics} {Simulations}.
  \emph{Journal of Chemical Information and Modeling} \textbf{2022}, Publisher:
  American Chemical Society\relax
\mciteBstWouldAddEndPuncttrue
\mciteSetBstMidEndSepPunct{\mcitedefaultmidpunct}
{\mcitedefaultendpunct}{\mcitedefaultseppunct}\relax
\EndOfBibitem
\bibitem[Liao \latin{et~al.}(2022)Liao, Nie, Unarta, Ericksen, and
  Tang]{liao_silico_2022}
Liao,~J.; Nie,~X.; Unarta,~I.~C.; Ericksen,~S.~S.; Tang,~W. In {Silico}
  {Modeling} and {Scoring} of {PROTAC}-{Mediated} {Ternary} {Complex} {Poses}.
  \emph{Journal of Medicinal Chemistry} \textbf{2022}, \emph{65}, 6116--6132,
  Publisher: American Chemical Society\relax
\mciteBstWouldAddEndPuncttrue
\mciteSetBstMidEndSepPunct{\mcitedefaultmidpunct}
{\mcitedefaultendpunct}{\mcitedefaultseppunct}\relax
\EndOfBibitem
\bibitem[Niesen \latin{et~al.}(2017)Niesen, Wang, Lehn, and
  Miller]{niesen_structurally_2017}
Niesen,~M. J.~M.; Wang,~C.~Y.; Lehn,~R. C.~V.; Miller,~T.~F. Structurally
  detailed coarse-grained model for {Sec}-facilitated co-translational protein
  translocation and membrane integration. \emph{PLOS Computational Biology}
  \textbf{2017}, \emph{13}, e1005427, Publisher: Public Library of
  Science\relax
\mciteBstWouldAddEndPuncttrue
\mciteSetBstMidEndSepPunct{\mcitedefaultmidpunct}
{\mcitedefaultendpunct}{\mcitedefaultseppunct}\relax
\EndOfBibitem
\bibitem[Zhang and Miller(2012)Zhang, and Miller]{zhang_long-timescale_2012}
Zhang,~B.; Miller,~T.~F. Long-{Timescale} {Dynamics} and {Regulation} of
  {Sec}-{Facilitated} {Protein} {Translocation}. \emph{Cell Reports}
  \textbf{2012}, \emph{2}, 927--937, Publisher: Elsevier\relax
\mciteBstWouldAddEndPuncttrue
\mciteSetBstMidEndSepPunct{\mcitedefaultmidpunct}
{\mcitedefaultendpunct}{\mcitedefaultseppunct}\relax
\EndOfBibitem
\bibitem[Hanke \latin{et~al.}(2010)Hanke, Serr, Kreuzer, and
  Netz]{hanke_stretching_2010}
Hanke,~F.; Serr,~A.; Kreuzer,~H.~J.; Netz,~R.~R. Stretching single
  polypeptides: {The} effect of rotational constraints in the backbone.
  \emph{EPL (Europhysics Letters)} \textbf{2010}, \emph{92}, 53001, Publisher:
  IOP Publishing\relax
\mciteBstWouldAddEndPuncttrue
\mciteSetBstMidEndSepPunct{\mcitedefaultmidpunct}
{\mcitedefaultendpunct}{\mcitedefaultseppunct}\relax
\EndOfBibitem
\bibitem[Staple \latin{et~al.}(2008)Staple, Payne, Reddin, and
  Kreuzer]{staple_model_2008}
Staple,~D.~B.; Payne,~S.~H.; Reddin,~A. L.~C.; Kreuzer,~H.~J. Model for
  {Stretching} and {Unfolding} the {Giant} {Multidomain} {Muscle} {Protein}
  {Using} {Single}-{Molecule} {Force} {Spectroscopy}. \emph{Physical Review
  Letters} \textbf{2008}, \emph{101}, 248301, Publisher: American Physical
  Society\relax
\mciteBstWouldAddEndPuncttrue
\mciteSetBstMidEndSepPunct{\mcitedefaultmidpunct}
{\mcitedefaultendpunct}{\mcitedefaultseppunct}\relax
\EndOfBibitem
\bibitem[{Charmainne Cruje} and {Devika B Chithrani}(2014){Charmainne Cruje},
  and {Devika B Chithrani}]{charmainne_cruje_polyethylene_2014}
{Charmainne Cruje},; {Devika B Chithrani}, Polyethylene {Glycol} {Density} and
  {Length} {Affects} {Nanoparticle} {Uptake} by {Cancer} {Cells}. \emph{Journal
  of Nanomedicine Research} \textbf{2014}, \emph{Volume 1}, Publisher: MedCrave
  Publishing\relax
\mciteBstWouldAddEndPuncttrue
\mciteSetBstMidEndSepPunct{\mcitedefaultmidpunct}
{\mcitedefaultendpunct}{\mcitedefaultseppunct}\relax
\EndOfBibitem
\bibitem[Lezon \latin{et~al.}(2009)Lezon, Shrivastava, Yang, and
  Bahar]{lezon_elastic_2009}
Lezon,~T.~R.; Shrivastava,~I.~H.; Yang,~Z.; Bahar,~I. \emph{Handbook on
  {Biological} {Networks}}; World {Scientific} {Lecture} {Notes} in {Complex}
  {Systems} Volume 10; WORLD SCIENTIFIC, 2009; Vol. Volume 10; pp
  129--158\relax
\mciteBstWouldAddEndPuncttrue
\mciteSetBstMidEndSepPunct{\mcitedefaultmidpunct}
{\mcitedefaultendpunct}{\mcitedefaultseppunct}\relax
\EndOfBibitem
\bibitem[Ricardo Batista \latin{et~al.}(2010)Ricardo Batista,
  Herbert Robert, Maréchal, Ben Hamida-Rebaï, Geraldo Pascutti,
  Mascarello Bisch, and Perahia]{ricardobatista_consensus_2010}
Ricardo Batista,~P.; Herbert Robert,~C.; Maréchal,~J.-D.;
  Ben Hamida-Rebaï,~M.; Geraldo Pascutti,~P.; Mascarello Bisch,~P.;
  Perahia,~D. Consensus modes, a robust description of protein collective
  motions from multiple-minima normal mode analysis—application to the
  {HIV}-1 protease. \emph{Physical Chemistry Chemical Physics} \textbf{2010},
  \emph{12}, 2850--2859, Publisher: Royal Society of Chemistry\relax
\mciteBstWouldAddEndPuncttrue
\mciteSetBstMidEndSepPunct{\mcitedefaultmidpunct}
{\mcitedefaultendpunct}{\mcitedefaultseppunct}\relax
\EndOfBibitem
\bibitem[Pohorille \latin{et~al.}(2010)Pohorille, Jarzynski, and
  Chipot]{pohorille_good_2010}
Pohorille,~A.; Jarzynski,~C.; Chipot,~C. Good {Practices} in {Free}-{Energy}
  {Calculations}. \emph{The Journal of Physical Chemistry B} \textbf{2010},
  \emph{114}, 10235--10253, Publisher: American Chemical Society\relax
\mciteBstWouldAddEndPuncttrue
\mciteSetBstMidEndSepPunct{\mcitedefaultmidpunct}
{\mcitedefaultendpunct}{\mcitedefaultseppunct}\relax
\EndOfBibitem
\bibitem[Klimovich \latin{et~al.}(2015)Klimovich, Shirts, and
  Mobley]{klimovich_guidelines_2015}
Klimovich,~P.~V.; Shirts,~M.~R.; Mobley,~D.~L. Guidelines for the analysis of
  free energy calculations. \emph{Journal of Computer-Aided Molecular Design}
  \textbf{2015}, \emph{29}, 397--411\relax
\mciteBstWouldAddEndPuncttrue
\mciteSetBstMidEndSepPunct{\mcitedefaultmidpunct}
{\mcitedefaultendpunct}{\mcitedefaultseppunct}\relax
\EndOfBibitem
\bibitem[Mey \latin{et~al.}(2020)Mey, Allen, Macdonald, Chodera, Kuhn, Michel,
  Mobley, Naden, Prasad, Rizzi, Scheen, Shirts, Tresadern, and
  Xu]{mey_best_2020}
Mey,~A. S. J.~S.; Allen,~B.; Macdonald,~H. E.~B.; Chodera,~J.~D.; Kuhn,~M.;
  Michel,~J.; Mobley,~D.~L.; Naden,~L.~N.; Prasad,~S.; Rizzi,~A. \latin{et~al.}
   Best {Practices} for {Alchemical} {Free} {Energy} {Calculations}.
  \emph{Living Journal of Computational Molecular Science} \textbf{2020},
  \emph{2}, arXiv: 2008.03067\relax
\mciteBstWouldAddEndPuncttrue
\mciteSetBstMidEndSepPunct{\mcitedefaultmidpunct}
{\mcitedefaultendpunct}{\mcitedefaultseppunct}\relax
\EndOfBibitem
\bibitem[Kirkwood(1935)]{kirkwood_statistical_1935}
Kirkwood,~J.~G. Statistical {Mechanics} of {Fluid} {Mixtures}. \emph{The
  Journal of Chemical Physics} \textbf{1935}, \emph{3}, 300--313, Publisher:
  American Institute of Physics\relax
\mciteBstWouldAddEndPuncttrue
\mciteSetBstMidEndSepPunct{\mcitedefaultmidpunct}
{\mcitedefaultendpunct}{\mcitedefaultseppunct}\relax
\EndOfBibitem
\bibitem[Bennett(1976)]{bennett_efficient_1976}
Bennett,~C.~H. Efficient estimation of free energy differences from {Monte}
  {Carlo} data. \emph{Journal of Computational Physics} \textbf{1976},
  \emph{22}, 245--268\relax
\mciteBstWouldAddEndPuncttrue
\mciteSetBstMidEndSepPunct{\mcitedefaultmidpunct}
{\mcitedefaultendpunct}{\mcitedefaultseppunct}\relax
\EndOfBibitem
\bibitem[Shirts and Chodera(2008)Shirts, and
  Chodera]{shirts_statistically_2008}
Shirts,~M.~R.; Chodera,~J.~D. Statistically optimal analysis of samples from
  multiple equilibrium states. \emph{The Journal of Chemical Physics}
  \textbf{2008}, \emph{129}\relax
\mciteBstWouldAddEndPuncttrue
\mciteSetBstMidEndSepPunct{\mcitedefaultmidpunct}
{\mcitedefaultendpunct}{\mcitedefaultseppunct}\relax
\EndOfBibitem
\bibitem[Clark \latin{et~al.}(2017)Clark, Gindin, Zhang, Wang, Abel, Murret,
  Xu, Bao, Lu, Zhou, Kwong, Shapiro, Honig, and Friesner]{clark_free_2017}
Clark,~A.~J.; Gindin,~T.; Zhang,~B.; Wang,~L.; Abel,~R.; Murret,~C.~S.; Xu,~F.;
  Bao,~A.; Lu,~N.~J.; Zhou,~T. \latin{et~al.}  Free {Energy} {Perturbation}
  {Calculation} of {Relative} {Binding} {Free} {Energy} between {Broadly}
  {Neutralizing} {Antibodies} and the gp120 {Glycoprotein} of {HIV}-1.
  \emph{Journal of Molecular Biology} \textbf{2017}, \emph{429}, 930--947\relax
\mciteBstWouldAddEndPuncttrue
\mciteSetBstMidEndSepPunct{\mcitedefaultmidpunct}
{\mcitedefaultendpunct}{\mcitedefaultseppunct}\relax
\EndOfBibitem
\bibitem[Clark \latin{et~al.}(2019)Clark, Negron, Hauser, Sun, Wang, Abel, and
  Friesner]{clark_relative_2019}
Clark,~A.~J.; Negron,~C.; Hauser,~K.; Sun,~M.; Wang,~L.; Abel,~R.;
  Friesner,~R.~A. Relative {Binding} {Affinity} {Prediction} of
  {Charge}-{Changing} {Sequence} {Mutations} with {FEP} in
  {Protein}–{Protein} {Interfaces}. \emph{Journal of Molecular Biology}
  \textbf{2019}, \emph{431}, 1481--1493\relax
\mciteBstWouldAddEndPuncttrue
\mciteSetBstMidEndSepPunct{\mcitedefaultmidpunct}
{\mcitedefaultendpunct}{\mcitedefaultseppunct}\relax
\EndOfBibitem
\bibitem[Patel \latin{et~al.}(2021)Patel, Patel, and
  Ytreberg]{patel_implementing_2021}
Patel,~D.; Patel,~J.~S.; Ytreberg,~F.~M. Implementing and {Assessing} an
  {Alchemical} {Method} for {Calculating} {Protein}–{Protein} {Binding}
  {Free} {Energy}. \emph{Journal of Chemical Theory and Computation}
  \textbf{2021}, \emph{17}, 2457--2464, Publisher: American Chemical
  Society\relax
\mciteBstWouldAddEndPuncttrue
\mciteSetBstMidEndSepPunct{\mcitedefaultmidpunct}
{\mcitedefaultendpunct}{\mcitedefaultseppunct}\relax
\EndOfBibitem
\bibitem[La~Serra \latin{et~al.}(2022)La~Serra, Vidossich, Acquistapace,
  Ganesan, and De~Vivo]{la_serra_alchemical_2022}
La~Serra,~M.~A.; Vidossich,~P.; Acquistapace,~I.; Ganesan,~A.~K.; De~Vivo,~M.
  Alchemical {Free} {Energy} {Calculations} to {Investigate}
  {Protein}–{Protein} {Interactions}: the {Case} of the {CDC42}/{PAK1}
  {Complex}. \emph{Journal of Chemical Information and Modeling} \textbf{2022},
  \emph{62}, 3023--3033, Publisher: American Chemical Society\relax
\mciteBstWouldAddEndPuncttrue
\mciteSetBstMidEndSepPunct{\mcitedefaultmidpunct}
{\mcitedefaultendpunct}{\mcitedefaultseppunct}\relax
\EndOfBibitem
\bibitem[Nandigrami \latin{et~al.}(2022)Nandigrami, Szczepaniak, Boughter,
  Dehez, Chipot, and Roux]{nandigrami_computational_2022}
Nandigrami,~P.; Szczepaniak,~F.; Boughter,~C.~T.; Dehez,~F.; Chipot,~C.;
  Roux,~B. Computational {Assessment} of {Protein}–{Protein} {Binding}
  {Specificity} within a {Family} of {Synaptic} {Surface} {Receptors}.
  \emph{The Journal of Physical Chemistry B} \textbf{2022}, Publisher: American
  Chemical Society\relax
\mciteBstWouldAddEndPuncttrue
\mciteSetBstMidEndSepPunct{\mcitedefaultmidpunct}
{\mcitedefaultendpunct}{\mcitedefaultseppunct}\relax
\EndOfBibitem
\bibitem[Sun \latin{et~al.}(2021)Sun, Ramaswamy, Levy, and
  Deng]{sun_computational_2021}
Sun,~Q.; Ramaswamy,~V. S.~K.; Levy,~R.; Deng,~N. Computational design of small
  molecular modulators of protein–protein interactions with a novel
  thermodynamic cycle: {Allosteric} inhibitors of {HIV}-1 integrase.
  \emph{Protein Science} \textbf{2021}, \emph{30}, 438--447, \_eprint:
  https://onlinelibrary.wiley.com/doi/pdf/10.1002/pro.4004\relax
\mciteBstWouldAddEndPuncttrue
\mciteSetBstMidEndSepPunct{\mcitedefaultmidpunct}
{\mcitedefaultendpunct}{\mcitedefaultseppunct}\relax
\EndOfBibitem
\bibitem[Testa \latin{et~al.}(2020)Testa, Hughes, Lucas, Wright, and
  Ciulli]{testa_structure-based_2020}
Testa,~A.; Hughes,~S.~J.; Lucas,~X.; Wright,~J.~E.; Ciulli,~A.
  Structure-{Based} {Design} of a {Macrocyclic} {PROTAC}. \emph{Angewandte
  Chemie International Edition} \textbf{2020}, \emph{59}, 1727--1734, \_eprint:
  https://onlinelibrary.wiley.com/doi/pdf/10.1002/anie.201914396\relax
\mciteBstWouldAddEndPuncttrue
\mciteSetBstMidEndSepPunct{\mcitedefaultmidpunct}
{\mcitedefaultendpunct}{\mcitedefaultseppunct}\relax
\EndOfBibitem
\bibitem[Hendrick \latin{et~al.}(2022)Hendrick, Jorgensen, Chaudhry,
  Strambeanu, Brazeau, Schiffer, Shi, Venable, and
  Wolkenberg]{hendrick_direct--biology_2022}
Hendrick,~C.~E.; Jorgensen,~J.~R.; Chaudhry,~C.; Strambeanu,~I.~I.;
  Brazeau,~J.-F.; Schiffer,~J.; Shi,~Z.; Venable,~J.~D.; Wolkenberg,~S.~E.
  Direct-to-{Biology} {Accelerates} {PROTAC} {Synthesis} and the {Evaluation}
  of {Linker} {Effects} on {Permeability} and {Degradation}. \emph{ACS
  Medicinal Chemistry Letters} \textbf{2022}, \emph{13}, 1182--1190, Publisher:
  American Chemical Society\relax
\mciteBstWouldAddEndPuncttrue
\mciteSetBstMidEndSepPunct{\mcitedefaultmidpunct}
{\mcitedefaultendpunct}{\mcitedefaultseppunct}\relax
\EndOfBibitem
\bibitem[J. Maple \latin{et~al.}(2019)J. Maple, Clayden, Baron, Stacey, and
  Felix]{jmaple_developing_2019}
J. Maple,~H.; Clayden,~N.; Baron,~A.; Stacey,~C.; Felix,~R. Developing
  degraders: principles and perspectives on design and chemical space.
  \emph{MedChemComm} \textbf{2019}, \emph{10}, 1755--1764, Publisher: Royal
  Society of Chemistry\relax
\mciteBstWouldAddEndPuncttrue
\mciteSetBstMidEndSepPunct{\mcitedefaultmidpunct}
{\mcitedefaultendpunct}{\mcitedefaultseppunct}\relax
\EndOfBibitem
\bibitem[Ermondi \latin{et~al.}(2020)Ermondi, Vallaro, and
  Caron]{ermondi_degraders_2020}
Ermondi,~G.; Vallaro,~M.; Caron,~G. Degraders early developability assessment:
  face-to-face with molecular properties. \emph{Drug Discovery Today}
  \textbf{2020}, \emph{25}, 1585--1591\relax
\mciteBstWouldAddEndPuncttrue
\mciteSetBstMidEndSepPunct{\mcitedefaultmidpunct}
{\mcitedefaultendpunct}{\mcitedefaultseppunct}\relax
\EndOfBibitem
\bibitem[Gopalsamy(2022)]{gopalsamy_selectivity_2022}
Gopalsamy,~A. Selectivity through {Targeted} {Protein} {Degradation} ({TPD}).
  \emph{Journal of Medicinal Chemistry} \textbf{2022}, \emph{65}, 8113--8126,
  Publisher: American Chemical Society\relax
\mciteBstWouldAddEndPuncttrue
\mciteSetBstMidEndSepPunct{\mcitedefaultmidpunct}
{\mcitedefaultendpunct}{\mcitedefaultseppunct}\relax
\EndOfBibitem
\bibitem[Towns \latin{et~al.}(2014)Towns, Cockerill, Dahan, Foster, Gaither,
  Grimshaw, Hazlewood, Lathrop, Lifka, Peterson, Roskies, Scott, and
  Wilkins-Diehr]{towns_xsede_2014}
Towns,~J.; Cockerill,~T.; Dahan,~M.; Foster,~I.; Gaither,~K.; Grimshaw,~A.;
  Hazlewood,~V.; Lathrop,~S.; Lifka,~D.; Peterson,~G.~D. \latin{et~al.}
  {XSEDE}: {Accelerating} {Scientific} {Discovery}. \emph{Computing in Science
  \& Engineering} \textbf{2014}, \emph{16}, 62--74, Conference Name: Computing
  in Science \& Engineering\relax
\mciteBstWouldAddEndPuncttrue
\mciteSetBstMidEndSepPunct{\mcitedefaultmidpunct}
{\mcitedefaultendpunct}{\mcitedefaultseppunct}\relax
\EndOfBibitem
\end{mcitethebibliography}


\providecommand{\latin}[1]{#1}
\makeatletter
\providecommand{\doi}
  {\begingroup\let\do\@makeother\dospecials
  \catcode`\{=1 \catcode`\}=2 \doi@aux}
\providecommand{\doi@aux}[1]{\endgroup\texttt{#1}}
\makeatother
\providecommand*\mcitethebibliography{\thebibliography}
\csname @ifundefined\endcsname{endmcitethebibliography}
  {\let\endmcitethebibliography\endthebibliography}{}
\begin{mcitethebibliography}{15}
\providecommand*\natexlab[1]{#1}
\providecommand*\mciteSetBstSublistMode[1]{}
\providecommand*\mciteSetBstMaxWidthForm[2]{}
\providecommand*\mciteBstWouldAddEndPuncttrue
  {\def\EndOfBibitem{\unskip.}}
\providecommand*\mciteBstWouldAddEndPunctfalse
  {\let\EndOfBibitem\relax}
\providecommand*\mciteSetBstMidEndSepPunct[3]{}
\providecommand*\mciteSetBstSublistLabelBeginEnd[3]{}
\providecommand*\EndOfBibitem{}
\mciteSetBstSublistMode{f}
\mciteSetBstMaxWidthForm{subitem}{(\alph{mcitesubitemcount})}
\mciteSetBstSublistLabelBeginEnd
  {\mcitemaxwidthsubitemform\space}
  {\relax}
  {\relax}

\bibitem[Gadd \latin{et~al.}(2017)Gadd, Testa, Lucas, Chan, Chen, Lamont,
  Zengerle, and Ciulli]{gadd_structural_2017}
Gadd,~M.~S.; Testa,~A.; Lucas,~X.; Chan,~K.-H.; Chen,~W.; Lamont,~D.~J.;
  Zengerle,~M.; Ciulli,~A. Structural basis of {PROTAC} cooperative recognition
  for selective protein degradation. \emph{Nature Chemical Biology}
  \textbf{2017}, \emph{13}, 514--521, Number: 5 Publisher: Nature Publishing
  Group\relax
\mciteBstWouldAddEndPuncttrue
\mciteSetBstMidEndSepPunct{\mcitedefaultmidpunct}
{\mcitedefaultendpunct}{\mcitedefaultseppunct}\relax
\EndOfBibitem
\bibitem[Nowak \latin{et~al.}(2018)Nowak, DeAngelo, Buckley, He, Donovan, An,
  Safaee, Jedrychowski, Ponthier, Ishoey, Zhang, Mancias, Gray, Bradner, and
  Fischer]{nowak_plasticity_2018}
Nowak,~R.~P.; DeAngelo,~S.~L.; Buckley,~D.; He,~Z.; Donovan,~K.~A.; An,~J.;
  Safaee,~N.; Jedrychowski,~M.~P.; Ponthier,~C.~M.; Ishoey,~M.; Zhang,~T.;
  Mancias,~J.~D.; Gray,~N.~S.; Bradner,~J.~E.; Fischer,~E.~S. Plasticity in
  binding confers selectivity in ligand-induced protein degradation.
  \emph{Nature Chemical Biology} \textbf{2018}, \emph{14}, 706--714, Number: 7
  Publisher: Nature Publishing Group\relax
\mciteBstWouldAddEndPuncttrue
\mciteSetBstMidEndSepPunct{\mcitedefaultmidpunct}
{\mcitedefaultendpunct}{\mcitedefaultseppunct}\relax
\EndOfBibitem
\bibitem[Schiemer \latin{et~al.}(2021)Schiemer, Horst, Meng, Montgomery, Xu,
  Feng, Borzilleri, Uccello, Leverett, Brown, Che, Brown, Hayward, Gilbert,
  Noe, and Calabrese]{schiemer_snapshots_2021}
Schiemer,~J.; Horst,~R.; Meng,~Y.; Montgomery,~J.~I.; Xu,~Y.; Feng,~X.;
  Borzilleri,~K.; Uccello,~D.~P.; Leverett,~C.; Brown,~S.; Che,~Y.;
  Brown,~M.~F.; Hayward,~M.~M.; Gilbert,~A.~M.; Noe,~M.~C.; Calabrese,~M.~F.
  Snapshots and ensembles of {BTK} and {cIAP1} protein degrader ternary
  complexes. \emph{Nature Chemical Biology} \textbf{2021}, \emph{17}, 152--160,
  Number: 2 Publisher: Nature Publishing Group\relax
\mciteBstWouldAddEndPuncttrue
\mciteSetBstMidEndSepPunct{\mcitedefaultmidpunct}
{\mcitedefaultendpunct}{\mcitedefaultseppunct}\relax
\EndOfBibitem
\bibitem[Madhavi~Sastry \latin{et~al.}(2013)Madhavi~Sastry, Adzhigirey, Day,
  Annabhimoju, and Sherman]{madhavi_sastry_protein_2013}
Madhavi~Sastry,~G.; Adzhigirey,~M.; Day,~T.; Annabhimoju,~R.; Sherman,~W.
  Protein and ligand preparation: parameters, protocols, and influence on
  virtual screening enrichments. \emph{Journal of Computer-Aided Molecular
  Design} \textbf{2013}, \emph{27}, 221--234\relax
\mciteBstWouldAddEndPuncttrue
\mciteSetBstMidEndSepPunct{\mcitedefaultmidpunct}
{\mcitedefaultendpunct}{\mcitedefaultseppunct}\relax
\EndOfBibitem
\bibitem[Niesen \latin{et~al.}(2017)Niesen, Wang, Lehn, and
  Miller]{niesen_structurally_2017}
Niesen,~M. J.~M.; Wang,~C.~Y.; Lehn,~R. C.~V.; Miller,~T.~F. Structurally
  detailed coarse-grained model for {Sec}-facilitated co-translational protein
  translocation and membrane integration. \emph{PLOS Computational Biology}
  \textbf{2017}, \emph{13}, e1005427, Publisher: Public Library of
  Science\relax
\mciteBstWouldAddEndPuncttrue
\mciteSetBstMidEndSepPunct{\mcitedefaultmidpunct}
{\mcitedefaultendpunct}{\mcitedefaultseppunct}\relax
\EndOfBibitem
\bibitem[Liwo \latin{et~al.}(1997)Liwo, Ołdziej, Pincus, Wawak, Rackovsky, and
  Scheraga]{liwo_united-residue_1997}
Liwo,~A.; Ołdziej,~S.; Pincus,~M.~R.; Wawak,~R.~J.; Rackovsky,~S.;
  Scheraga,~H.~A. A united-residue force field for off-lattice
  protein-structure simulations. {I}. {Functional} forms and parameters of
  long-range side-chain interaction potentials from protein crystal data.
  \emph{Journal of Computational Chemistry} \textbf{1997}, \emph{18},
  849--873\relax
\mciteBstWouldAddEndPuncttrue
\mciteSetBstMidEndSepPunct{\mcitedefaultmidpunct}
{\mcitedefaultendpunct}{\mcitedefaultseppunct}\relax
\EndOfBibitem
\bibitem[Lezon \latin{et~al.}(2009)Lezon, Shrivastava, Yang, and
  Bahar]{lezon_elastic_2009}
Lezon,~T.~R.; Shrivastava,~I.~H.; Yang,~Z.; Bahar,~I. \emph{Handbook on
  {Biological} {Networks}}; World {Scientific} {Lecture} {Notes} in {Complex}
  {Systems} Volume 10; WORLD SCIENTIFIC, 2009; Vol. Volume 10; pp
  129--158\relax
\mciteBstWouldAddEndPuncttrue
\mciteSetBstMidEndSepPunct{\mcitedefaultmidpunct}
{\mcitedefaultendpunct}{\mcitedefaultseppunct}\relax
\EndOfBibitem
\bibitem[Ricardo Batista \latin{et~al.}(2010)Ricardo Batista,
  Herbert Robert, Maréchal, Ben Hamida-Rebaï, Geraldo Pascutti,
  Mascarello Bisch, and Perahia]{ricardobatista_consensus_2010}
Ricardo Batista,~P.; Herbert Robert,~C.; Maréchal,~J.-D.;
  Ben Hamida-Rebaï,~M.; Geraldo Pascutti,~P.; Mascarello Bisch,~P.;
  Perahia,~D. Consensus modes, a robust description of protein collective
  motions from multiple-minima normal mode analysis—application to the
  {HIV}-1 protease. \emph{Physical Chemistry Chemical Physics} \textbf{2010},
  \emph{12}, 2850--2859, Publisher: Royal Society of Chemistry\relax
\mciteBstWouldAddEndPuncttrue
\mciteSetBstMidEndSepPunct{\mcitedefaultmidpunct}
{\mcitedefaultendpunct}{\mcitedefaultseppunct}\relax
\EndOfBibitem
\bibitem[Periole \latin{et~al.}(2009)Periole, Cavalli, Marrink, and
  Ceruso]{periole_combining_2009}
Periole,~X.; Cavalli,~M.; Marrink,~S.-J.; Ceruso,~M.~A. Combining an {Elastic}
  {Network} {With} a {Coarse}-{Grained} {Molecular} {Force} {Field}:
  {Structure}, {Dynamics}, and {Intermolecular} {Recognition}. \emph{Journal of
  Chemical Theory and Computation} \textbf{2009}, \emph{5}, 2531--2543,
  Publisher: American Chemical Society\relax
\mciteBstWouldAddEndPuncttrue
\mciteSetBstMidEndSepPunct{\mcitedefaultmidpunct}
{\mcitedefaultendpunct}{\mcitedefaultseppunct}\relax
\EndOfBibitem
\bibitem[Sievers \latin{et~al.}(2018)Sievers, Petzold, Bunker, Renneville,
  Słabicki, Liddicoat, Abdulrahman, Mikkelsen, Ebert, and
  Thomä]{sievers_defining_2018}
Sievers,~Q.~L.; Petzold,~G.; Bunker,~R.~D.; Renneville,~A.; Słabicki,~M.;
  Liddicoat,~B.~J.; Abdulrahman,~W.; Mikkelsen,~T.; Ebert,~B.~L.; Thomä,~N.~H.
  Defining the human {C2H2} zinc finger degrome targeted by thalidomide analogs
  through {CRBN}. \emph{Science} \textbf{2018}, \emph{362}, eaat0572,
  Publisher: American Association for the Advancement of Science\relax
\mciteBstWouldAddEndPuncttrue
\mciteSetBstMidEndSepPunct{\mcitedefaultmidpunct}
{\mcitedefaultendpunct}{\mcitedefaultseppunct}\relax
\EndOfBibitem
\bibitem[Pohorille \latin{et~al.}(2010)Pohorille, Jarzynski, and
  Chipot]{pohorille_good_2010}
Pohorille,~A.; Jarzynski,~C.; Chipot,~C. Good {Practices} in {Free}-{Energy}
  {Calculations}. \emph{The Journal of Physical Chemistry B} \textbf{2010},
  \emph{114}, 10235--10253, Publisher: American Chemical Society\relax
\mciteBstWouldAddEndPuncttrue
\mciteSetBstMidEndSepPunct{\mcitedefaultmidpunct}
{\mcitedefaultendpunct}{\mcitedefaultseppunct}\relax
\EndOfBibitem
\bibitem[Bennett(1976)]{bennett_efficient_1976}
Bennett,~C.~H. Efficient estimation of free energy differences from {Monte}
  {Carlo} data. \emph{Journal of Computational Physics} \textbf{1976},
  \emph{22}, 245--268\relax
\mciteBstWouldAddEndPuncttrue
\mciteSetBstMidEndSepPunct{\mcitedefaultmidpunct}
{\mcitedefaultendpunct}{\mcitedefaultseppunct}\relax
\EndOfBibitem
\bibitem[Klimovich \latin{et~al.}(2015)Klimovich, Shirts, and
  Mobley]{klimovich_guidelines_2015}
Klimovich,~P.~V.; Shirts,~M.~R.; Mobley,~D.~L. Guidelines for the analysis of
  free energy calculations. \emph{Journal of Computer-Aided Molecular Design}
  \textbf{2015}, \emph{29}, 397--411\relax
\mciteBstWouldAddEndPuncttrue
\mciteSetBstMidEndSepPunct{\mcitedefaultmidpunct}
{\mcitedefaultendpunct}{\mcitedefaultseppunct}\relax
\EndOfBibitem
\bibitem[Wyce \latin{et~al.}(2013)Wyce, Ganji, Smitheman, Chung, Korenchuk,
  Bai, Barbash, Le, Craggs, McCabe, Kennedy-Wilson, Sanchez, Gosmini, Parr,
  McHugh, Dhanak, Prinjha, Auger, and Tummino]{wyce_bet_2013}
Wyce,~A.; Ganji,~G.; Smitheman,~K.~N.; Chung,~C.-w.; Korenchuk,~S.; Bai,~Y.;
  Barbash,~O.; Le,~B.; Craggs,~P.~D.; McCabe,~M.~T.; Kennedy-Wilson,~K.~M.;
  Sanchez,~L.~V.; Gosmini,~R.~L.; Parr,~N.; McHugh,~C.~F.; Dhanak,~D.;
  Prinjha,~R.~K.; Auger,~K.~R.; Tummino,~P.~J. {BET} {Inhibition} {Silences}
  {Expression} of {MYCN} and {BCL2} and {Induces} {Cytotoxicity} in
  {Neuroblastoma} {Tumor} {Models}. \emph{PLOS ONE} \textbf{2013}, \emph{8},
  e72967, Publisher: Public Library of Science\relax
\mciteBstWouldAddEndPuncttrue
\mciteSetBstMidEndSepPunct{\mcitedefaultmidpunct}
{\mcitedefaultendpunct}{\mcitedefaultseppunct}\relax
\EndOfBibitem
\end{mcitethebibliography}

\end{document}


\section{CGMD Forcefield}

The complete potential energy function for a ternary complex is
\begin{equation}
\begin{aligned}
U\p{\boldsymbol{x};\boldsymbol{b},\boldsymbol{q}} &= U_\text{ENM}\p{\boldsymbol{x}_E} + U_\text{ENM}\p{\boldsymbol{x}_T} 
+ U_\text{spring}\p{\boldsymbol{x}_P} + U_\text{WCA}\p{\boldsymbol{x}_P} \\
&+ U_\text{bind}\p{\boldsymbol{x}_P, \boldsymbol{x}_T; \boldsymbol{b}}
 + U_\text{bind}\p{\boldsymbol{x}_P, \boldsymbol{x}_E; \boldsymbol{b}}
   + U_\text{WCA}\p{\boldsymbol{x}_P, \boldsymbol{x}_T}
   + U_\text{WCA}\p{\boldsymbol{x}_P, \boldsymbol{x}_E}\\
&+ U_\text{WCA}\p{\boldsymbol{x}_E, \boldsymbol{x}_T}
 + U_\text{elec}\p{\boldsymbol{x}_E, \boldsymbol{x}_T;\boldsymbol{q}}
 + U_\text{LJ}\p{\boldsymbol{x}_E, \boldsymbol{x}_T;\epsilon_\text{LJ}} 
   \label{Upotential}
\end{aligned}
\end{equation}
where $\boldsymbol{x}_E$, $\boldsymbol{x}_T$, and $\boldsymbol{x}_P$ indicate the coordinates of the E3 ligase, the target protein, and
the PROTAC respectively, $\boldsymbol{q}$ represent the charges of protein beads, and $\boldsymbol{b}$ are indicators of whether protein beads are at the binding pocket or not. All PROTAC beads are modeled with 0 charge and no attraction to the proteins. All parameters and variables are defined using a length scale of the large bead ($\sigma=0.8$ nm) and an energy scale of $\epsilon=kT$ where $k$ is the Boltzmann constant and $T=310$ K.

\subsection{Internal energy terms}
Interactions within a protein are modeled by an elastic network model (ENM) such that every pair of beads within distance $R_c$ is connected by a harmonic spring:
\begin{equation}
\begin{aligned}
    U_\text{ENM}\p{\boldsymbol{x}} 
    = \sum_{\p{i,j} \in D} k_\text{spring}\p{\Delta x_{ij}-d_{ij}}^2 
    \label{Uenm}
\end{aligned}
\end{equation}
where $k_\text{spring}$ is the spring constant, $d_{ij}$ is the optimal distance between $x_i$ and $x_j$, and $D=\{\p{i,j} |d_{ij} < R_c\}$. The optimal distance between a pair of beads is its initial distance in the experimental structure. Experimental structures used in this work include VHL (\textcolor{black}{PDB: 5T35\cite{gadd_structural_2017}} chain D),  BRD4\textsuperscript{BD2} (\textcolor{black}{PDB: 5T35\cite{gadd_structural_2017}} chain A), CRBN (\textcolor{black}{PDB: 6BOY\cite{nowak_plasticity_2018}} chain B), and BTK (\textcolor{black}{PDB: 6W7O\cite{schiemer_snapshots_2021}} chain A), and Schrödinger Maestro \cite{madhavi_sastry_protein_2013} is used to fill in missing atoms and perform energy minimization before building the CG ENM. Additional details on the parameterization are described in a separate section below.

PROTAC is modeled as a linear molecule, where adjacent beads are connected by springs ($U_\text{spring}\p{\boldsymbol{x}_P}$) and non-adjacent beads are subjected to steric repulsions ($U_\text{WCA}\p{\boldsymbol{x}_P}$). 

\subsection{Interaction energy terms}
PROTAC-protein interactions consist of binding interactions modeled by springs between a binding moiety bead in the PROTAC and all beads in the corresponding binding pocket ($U_\text{bind}\p{\boldsymbol{x}_P, \boldsymbol{x}_T; \boldsymbol{b}}$ and $U_\text{bind}\p{\boldsymbol{x}_P, \boldsymbol{x}_E; \boldsymbol{b}}$ in eq.\eqref{Upotential}) and steric repulsions ($U_\text{WCA}\p{\boldsymbol{x}_P, \boldsymbol{x}_T}$ and $U_\text{WCA}\p{\boldsymbol{x}_P, \boldsymbol{x}_E}$) between the remaining parts of PROTAC and protein. Steric repulsions in intra-PROTAC, PROTAC-protein, and inter-protein interactions are all modeled by the Weeks-Chandler-Andersen (WCA) potential, a shifted and truncated version of Lennard-Jones (LJ) potential. 

\begin{figure}[htbp!]
\centering
\includegraphics[page=1,scale=0.48]{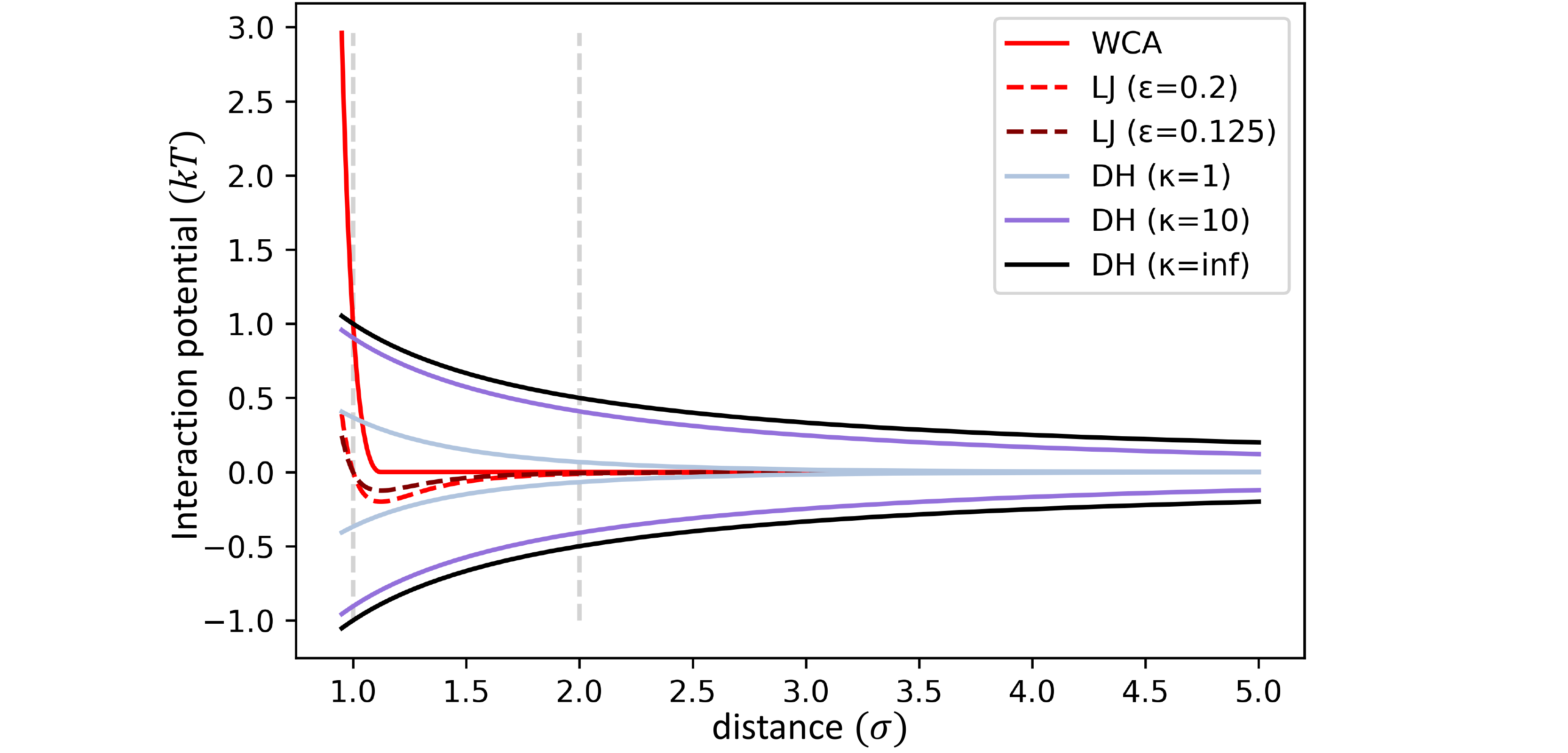}
\caption{The strengths of various interaction potentials are plotted over the distance between protein beads. The two vertical dashed grey lines bound the distance between 1 and 2 $\sigma$. The electrostatic potentials (DH) are plotted for beads with +1 and +1 charges or +1 and -1 charges.}
\label{fgr:potentials}
\end{figure}

Protein-protein interactions are captured by the steric repulsions ($U_\text{WCA}\p{\boldsymbol{x}_E, \boldsymbol{x}_T}$), and depending on the modeling purpose, electrostatics ($U_\text{elec}\p{\boldsymbol{x}_E, \boldsymbol{x}_T;\boldsymbol{q}}$) or nonspecific attractions ($U_\text{LJ}\p{\boldsymbol{x}_E, \boldsymbol{x}_T;\epsilon_\text{LJ}}$). The electrostatic interaction is modeled by a Debye-Hückel (DH) potential. The functional forms and parameterization of both potentials can be found in \cite{niesen_structurally_2017}. When reducing the screening of electrostatics between BRD4\textsuperscript{BD2} and VHL, the Debye length $\kappa$ is multiplied by 10. 
The solvent in our system is treated implicitly. Nonspecific attractions aimed at broadly including Van der Waals forces and hydrophobic interactions are modeled by LJ potentials. The strength of the attraction is kept under that of electrostatic interactions (Fig. \ref{fgr:potentials}). The well depth of LJ, $\epsilon_\text{LJ}$, is currently set to be the same for all pairs of beads for nonspecific attraction. For future efforts, minor modifications to the formula \cite{liwo_united-residue_1997} and parameterization of $\epsilon_\text{LJ}$ to depend on the Wimley-White hydrophobicity scale, for example, can capture more sequence-specific interactions such as the hydrophobic effects.

\subsection{Parameterization of ENM}
ENM is a model that represents the tertiary structure of a protein by connecting every pair of protein beads within a certain distance cutoff $R_c$ by a Hookean spring of spring constant $k_\text{spring}$. Despite the simplicity of its parameterization, 
slow modes in ENM can capture biologically significant conformational changes \cite{lezon_elastic_2009,ricardobatista_consensus_2010}. This structure-based model can also be used in combination with other physics-driven forcefields to model macromolecular complexes. Protein-protein associations and viral capsid assembly have both been successfully modeled by using Elnedyn, an ENM at the resolution of 1 residue per bead \cite{periole_combining_2009}, on top of the MARTINI CG forcefield. By fitting to atomistic simulations, Elnedyn preserves both structural properties and dynamics within each protein subunit for the CG simulations.

We follow a similar protocol \textcolor{black}{and fit our CG ENM parameters in eq.\eqref{Uenm} to Elnedyn simulations results.}
Three proteins -- IKZF1\textsuperscript{ZF2} (\textcolor{black}{PDB: 6H0F\cite{sievers_defining_2018}} chain C), BRD4\textsuperscript{BD1} (\textcolor{black}{PDB: 6BOY\cite{nowak_plasticity_2018}} chain C), and CRBN (\textcolor{black}{PDB: 6BOY\cite{nowak_plasticity_2018}} chain B) -- are chosen for the fitting to represent the range of protein sizes based on the publicly available crystal structures of PROTAC-mediated ternary complexes. 
Elnedyn is supported as an option in the MARTINI 2 CG forcefield\cite{periole_combining_2009}, and \textcolor{black}{we use the default parameters to generate Elnedyn simulations of these proteins with GROMACS version 5.0.7. Two equilibration stages were run, first at 1 fs timestep for 50 ps, and then at 10 fs timestep for 1 ns. Then, only the dynamics stage was used for fitting, which was run at 10 fs timestep for 40 ns. }
Four metrics are used to examine how well a particular combination of $k_\text{spring}$ and $R_c$ captures information in Elnedyn simulations: the difference of time-averaged root-mean-squared-deviation ($\Delta$RMSD), bead-averaged root-mean-squared-fluctuation ($\Delta$RMSF), Kullback–Leibler (KL) divergence of the RMSD distributions, and the root-mean-squared inner product of the principal components (RMSIP) of the trajectories. 

\begin{figure}[htbp!]
\centering
\includegraphics[page=2,scale=0.48]{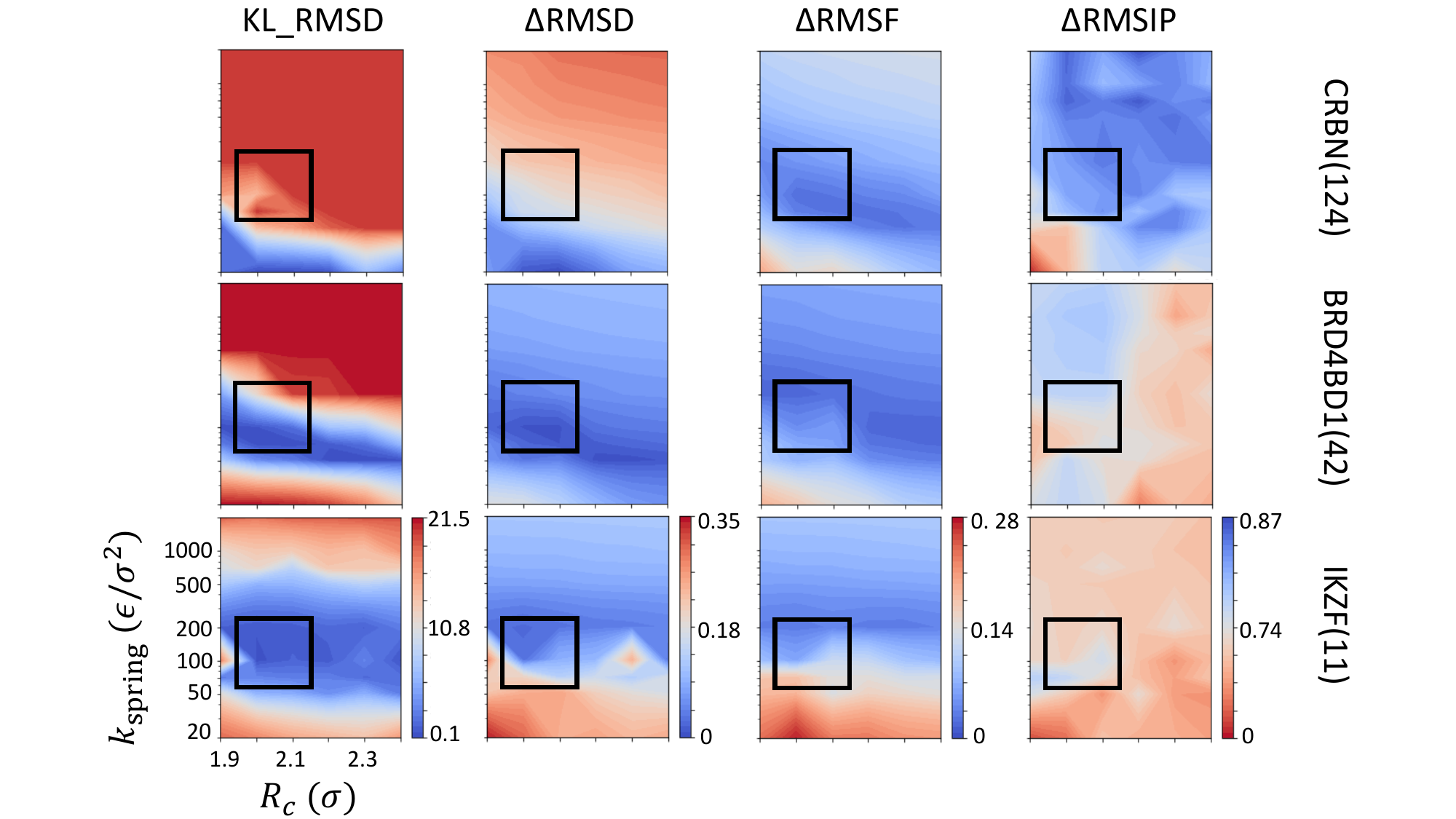}
\caption{Fitting results of ENM parameters arranged by proteins (rows) and evaluation metrics (columns). Numbers in parenthesis next to protein names are the number of CG beads. For each plot, blue regions indicate $k_\text{spring}$ and $R_c$ values that result in good fitting, and red regions indicate significant differences between our simulations and Elnedyn simulations. Each column shares the same colorbar range. In general, the boxed regions around $k_\text{spring}=100\epsilon/\sigma^2$ and $R_c=2.0\sigma$ has good fitting.}
\label{fgr:enm}
\end{figure}

Within a single metric, we usually observe a degeneracy within a certain region of $k_\text{spring}$ and $R_c$ values (Fig. \ref{fgr:enm}), and this was also observed in Elnedyn fitting to atomistic simulations \cite{periole_combining_2009}. This is because increasing either $k_\text{spring}$ or $R_c$ can increase the stiffness of a protein and, therefore, can compensate for each other to some extent. Nevertheless, despite the degeneracy, given the wide range of protein sizes, there is no single combination of $k_\text{spring}$ and $R_c$ values that works best for all three proteins. We chose $k_\text{spring}=100\epsilon/\sigma^2$ and $R_c=2.0\sigma$ as they are near the optimal degeneracy region under most metrics and consistent with the values of Elnedyn parameters ($k_\text{spring}=124.25\epsilon/\sigma^2$ and $R_c=1.125\sigma$). 
\textcolor{black}{
This combination of $k_\text{spring}$ and $R_c$ was selected without a global optimization function that combines all four metrics, and should be subjected to finer tuning if a specific system is of interest.}

\section{Analysis of alchemical free energy calculations} \label{sanity_section}
We perform various checks to address two common concerns in alchemical simulations: 1) are there sufficient intermediate states along the alchemical reaction pathway, and 2) are there sufficient samples from each state for accurate free energy calculations. The BTK-PROTAC (10)-CRBN complex is used as an example for the analysis below.

\begin{figure}[htbp!]
\centering
\includegraphics[page=3,scale=0.48]{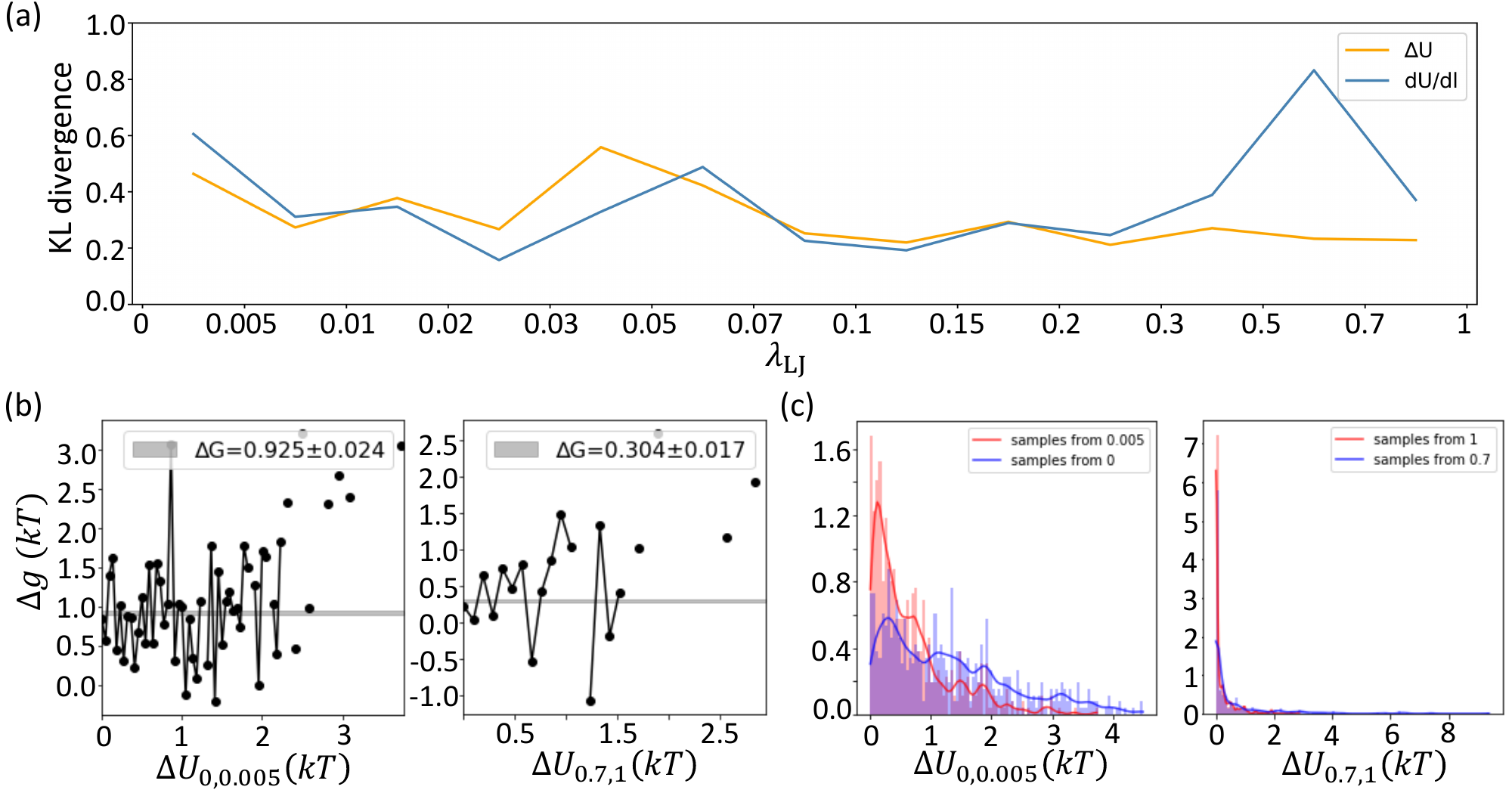}
\caption{Phase space overlap in calculating $\Delta G\textsuperscript{ternary(WCA)}$ for BTK-CRBN in Fig. 2. 
\textbf{(a)} Overlap of $\Delta U$ and $\partial U/\partial \lambda$ distributions between adjacent states are quantified by the KL divergence. 
\textbf{(b)} Example Bennett's overlapping plots for $\lambda_\text{LJ}=0, 0.005$ states (left) and $\lambda_\text{LJ}=0.7, 1$ states (right). The grey bands represent $\Delta G_{\lambda_{i},\lambda_{i+1}}$ $\pm 1$ std estimated using BAR.
\textbf{(c)} Example distributions of $\Delta U_{i,i+1}$ are shown with Gaussian smoothing (red and blue solid curves) for better visualization. }
\label{fgr:phasespace}
\end{figure}

We first validate that there are sufficient intermediate states for a converged estimation of $\Delta G\textsuperscript{ternary(WCA)}$. The convergence of free energy calculations depends on the overlap of the phase space, i.e. the distribution of sampled conformations, between neighboring states. Substantial overlap is achieved when the neighboring states are similar, which requires a fine spacing of the coupling parameter values. In practice, distributions of quantities such as $\Delta U$ and $\partial U/\partial \lambda$ that are directly involved in free energy estimations are often treated as proxies for the high-dimensional phase space \cite{pohorille_good_2010}. The similarity between distributions is quantified by KL divergence, where 0 indicates identical distributions and $\gg 1$ suggests concerning differences. Based on this metric, all neighboring states have substantial overlap, as the Kullback–Leibler (KL) divergence values of $\Delta U$ and of $\partial U/\partial \lambda$ distributions both stay below 1 (Fig. \ref{fgr:phasespace}a).

Bennett's overlapping histogram \cite{bennett_efficient_1976} provides another qualitative test for the overlap of $\Delta U$ distributions. The difference between $g_{\lambda_{i+1}}(\Delta U_{\lambda_{i},\lambda_{i+1}}) = P_{\lambda_{i}}(\Delta U_{\lambda_{i},\lambda_{i+1}}) + \p{1-C}\Delta U_{\lambda_{i},\lambda_{i+1}}$ and $g_{\lambda_{i}}(\Delta U_{\lambda_{i},\lambda_{i+1}}) = P_{\lambda_{i+1}}(\Delta U_{\lambda_{i},\lambda_{i+1}}) - C\Delta U_{\lambda_{i},\lambda_{i+1}}$ is plotted over $\Delta U_{\lambda_{i},\lambda_{i+1}}$ values, where $C$ is an arbitrary constant between 0 and 1 and $P_{\lambda_{i}}(\Delta U_{\lambda_{i},\lambda_{i+1}})$ and $P_{\lambda_{i}}(\Delta U_{\lambda_{i},\lambda_{i+1}})$ are the distributions of $\Delta U_{\lambda_{i},\lambda_{i+1}}$ obtained by sampling from neighboring alchemical states $\lambda_i$ and $\lambda_{i+1}$ respectively. Continuous oscillations of $g_{\lambda_{i+1}}(\Delta U_{\lambda_{i},\lambda_{i+1}}) - g_{\lambda_{i}}(\Delta U_{\lambda_{i},\lambda_{i+1}})$ around the estimated $\Delta G_{\lambda_{i},\lambda_{i+1}}$ over a range of $\Delta U_{\lambda_{i},\lambda_{i+1}}$ values suggests good overlap (Fig. \ref{fgr:phasespace}b) \cite{klimovich_guidelines_2015}. For states of higher $\lambda_\text{LJ}$ values, higher energetic penalty of steric repulsions prevents sampling over a wide range of $\Delta U$ values, but the KL divergence and visualization of the distributions (Fig. \ref{fgr:phasespace}a,c) both indicate the quality of the overlap.

\begin{figure}[H]
\centering
\includegraphics[page=4,scale=0.48]{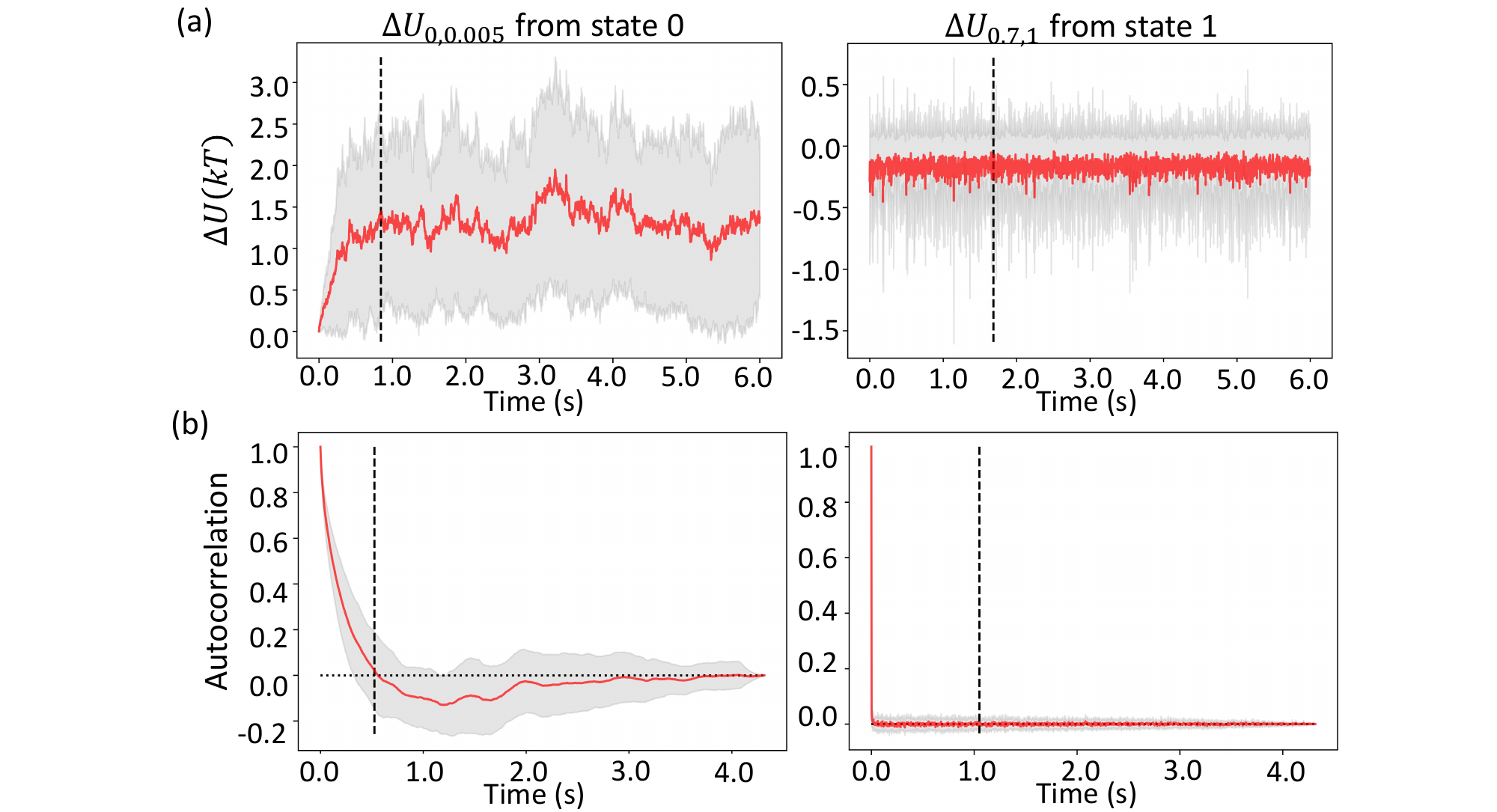}
\caption{Detecting equilibration and autocorrelation time in calculating $\Delta G\textsuperscript{ternary(WCA)}$ for BTK-CRBN in Fig. 2. \textbf{(a)} $\Delta U_{\lambda_{i},\lambda_{i+1}}$ over simulation time and \textbf{(b)} the autocorrelation of $\Delta U_{\lambda_{i},\lambda_{i+1}}$ from $\lambda_\text{LJ}=0$ (left) and $\lambda_\text{LJ}=1$ (right). The red curves and the shaded regions represent the average value $\pm 1$ standard deviation based on 64 independent trajectories. The vertical dashed lines in this example mark 0.9 s in (a) and 0.63 s in (b). The horizontal dotted lines in (b) mark the 0 autocorrelation value. }
\label{fgr:autocorr}
\end{figure}

Next, we examine sampling within each state. For each state, a simulation needs to be post-processed to discard the initial unequilibrated part and then subsampled to obtain de-correlated data for accurate uncertainty quantification of the free energy estimation. Thus, the length of the simulations is dictated by the equilibration time, autocorrelation time, and the number of de-correlated samples needed for converged estimations. We examine the values of $\Delta U$, $\partial U /\partial \lambda$, and other collective variables over the simulation time, which typically equilibrate after 0.9 s (Fig. \ref{fgr:autocorr}a). To find out the decorrelation time, we discard the initial 0.9 s of simulations and plot the autocorrelation functions of these variables over different time lags up to half of the simulation time to ensure that the autocorrelation is calculated from a sufficient number of samples. The autocorrelation times all plummet to 0 before 0.63 s (Fig. \ref{fgr:autocorr}b). Both equilibration time and decorrelation time are longer for simulations in lower value of $\lambda_\text{LJ}$ states that retain more memory of previously sampled configurations due to lower energetic costs. Currently, the equilibration and autocorrelation cutoffs depend on each system. For convenience, we used the same cutoffs for all $\lambda$ states. In the future, this can be customized for each state to maximize the number of samples, especially from states of high $\lambda$ values that requires less equilibration and decorrelation time (Fig. \ref{fgr:autocorr}b).

\begin{figure}[H]
\centering
\includegraphics[page=5,scale=0.48]{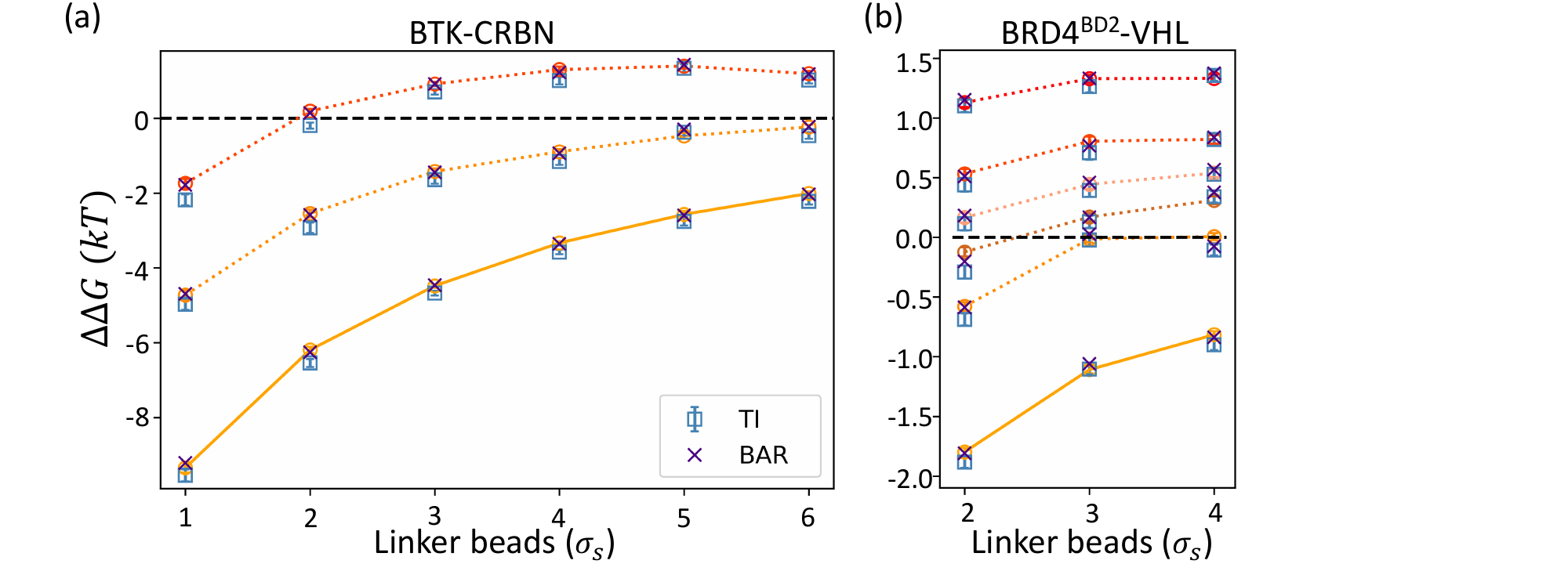}
\caption{$\Delta \Delta G$s calculated by TI and BAR are superimposed onto the MBAR results shown in Figure 3 to show that all three alchemical free energy calculation methods agree within noise.}
\label{fgr:ti_bar_fig3}
\end{figure}

\begin{figure}[H]
\centering
\includegraphics[page=6,scale=0.48]{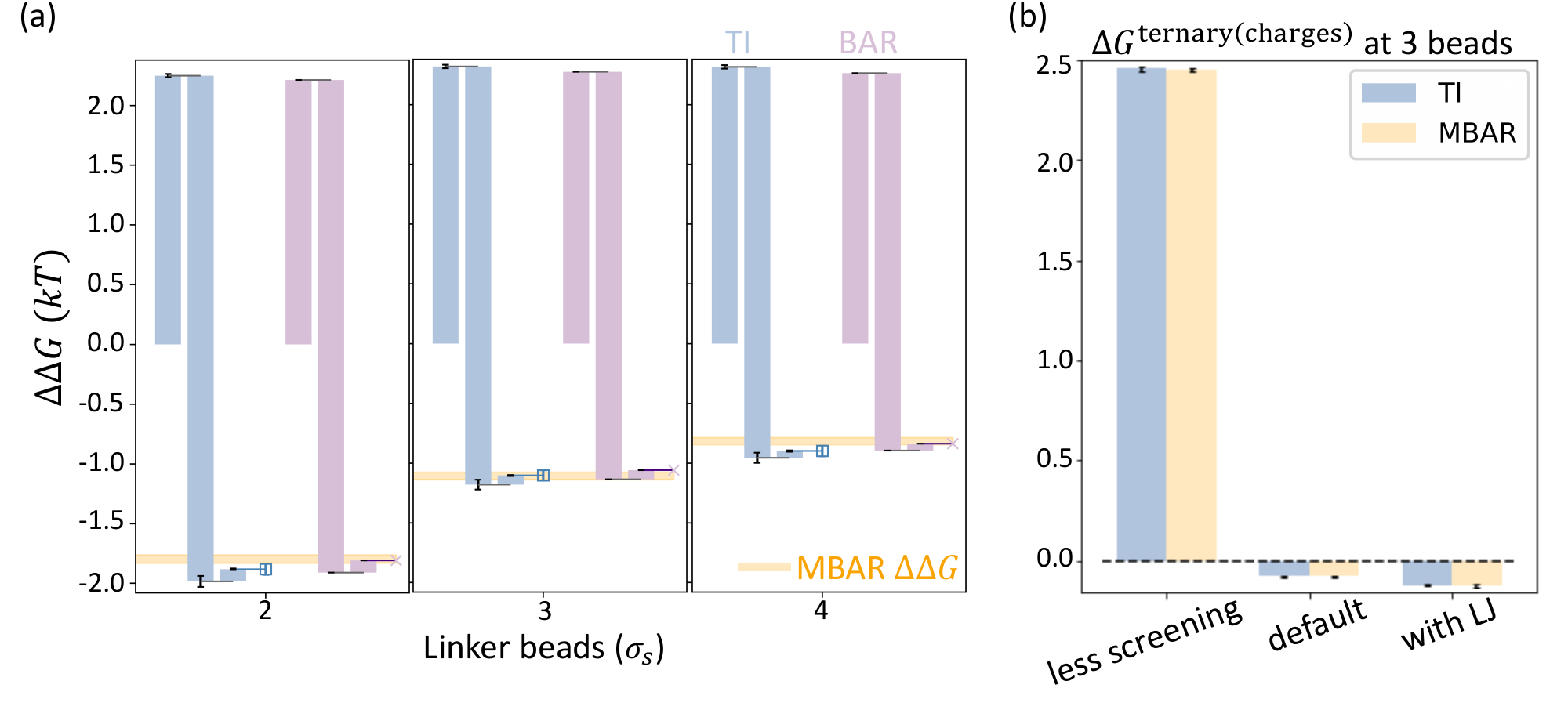}
\caption{$\Delta \Delta G$s calculated by TI and BAR agree with MBAR results shown in Figure 4 for the BRD4\textsuperscript{BD2}-VHL system modeled with protein charges included. \textbf{(a)} $\Delta \Delta G$s at each PROTAC linker length calculated by TI and BAR are broken down using waterfall plots similar to Figure 4b. In each triplet, columns from left to right correspond to $\Delta G\textsuperscript{binary}$, $- \Delta G\textsuperscript{ternary(other)}$, and $-\Delta G\textsuperscript{ternary(charges)}$. Columns are arranged cumulatively such that the end point of a triplet of columns represent the final $\Delta \Delta G$ value calculated by the corresponding method. MBAR $\Delta \Delta G$ values with $\pm 1$ standard deviation are shown as horizontal yellow bands for reference.
\textbf{(b)} TI and MBAR calculations of the electrostatic contribution to $\Delta \Delta G$ under different forcefield setups at the linker length of 3 beads agree with each other. Note that $\Delta G\textsuperscript{ternary(charges)}$ is shown here rather than $-\Delta G\textsuperscript{ternary(charges)}$ in panel \textbf{(a)}.}
\label{fgr:ti_bar_fig4}
\end{figure}

\begin{figure}[H]
\centering
\includegraphics[page=7,scale=0.48]{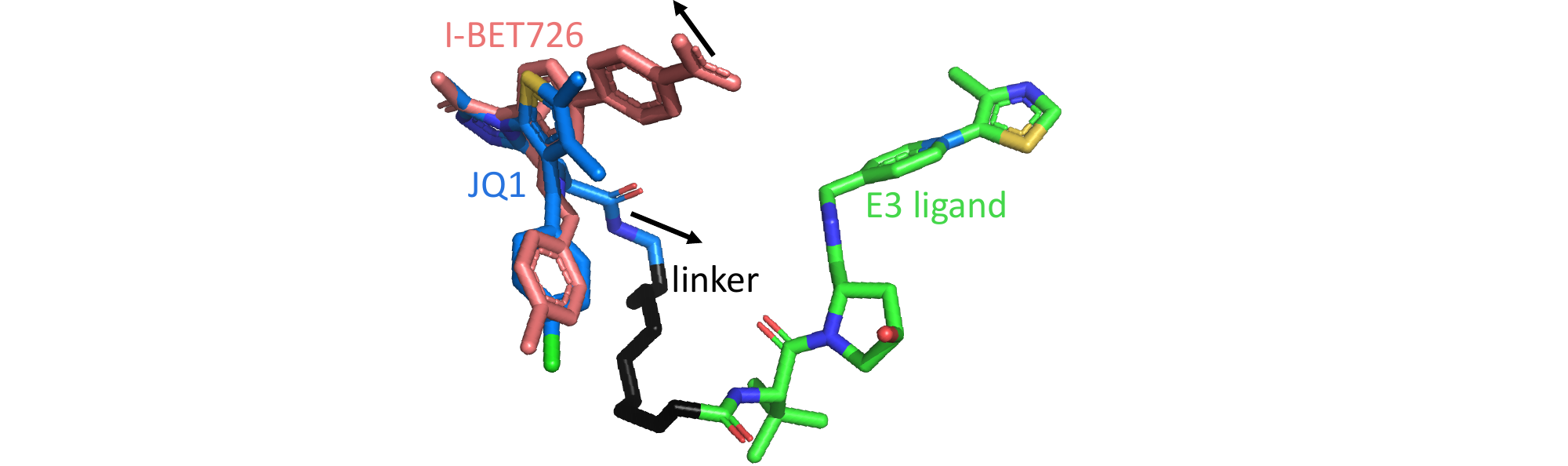}
\caption{The structure of MZ1, which is a PROTAC with linker length of 3 beads using a JQ1 warhead, extracted from the ternary crystal structure (\textcolor{black}{PDB: 5T35\cite{gadd_structural_2017}}) and the structure of I-BET726 warhead extracted from the crystal structure of a binary complex (\textcolor{black}{PDB: 4BJX\cite{wyce_bet_2013}}) are superimposed to highlight the difference in exit vectors (black arrows).}
\label{fgr:jq1_ibet726}
\end{figure}

\bibliography{references}